\documentclass[intlimits,twoside,a4paper]{article}

\usepackage[T2A]{fontenc} 
\usepackage[utf8]{inputenc}
\usepackage[eqsecnum]{cmpj3}

\newcommand{\fluc}{\delta\hat{G}}
\newcommand{\hh}{\mathrm{H}}
\newcommand{\HF}{\mathrm{HF}}
\newcommand{\cfluc}{\Delta n^\lambda}
\newcommand{\CC}{\mathbb{C}}

\issue{2022}{25}{2}{23401}
\doinumber{10.5488/CMP.25.23401}

\title[Quantum fluctuations approach to the nonequilibrium \emph{GW} approximation]%
{Quantum fluctuations approach to the nonequilibrium \emph{GW} approximation%
}

\author[E. Schroedter, J.~-P. Joost,
M. Bonitz]{E. Schroedter, J.~-P. Joost\orcid{0000-0003-2237-8162}, M. Bonitz\orcid{0000-0001-7911-0656}\thanks{Corresponding author: \email{bonitz@theo-physik.uni-kiel.de}}}
\address{Institut f\"ur Theoretische Physik und Astrophysik, 
Christian-Albrechts-Universit\"{a}t zu Kiel, D-24098 Kiel, Germany
}


\date{Received April 15, 2022, in final form June 03, 2022}

\begin{document}

\maketitle

\begin{abstract}
The quantum dynamics of fermionic or bosonic many-body systems following external excitation can be successfully studied using two-time nonequilibrium Green's functions (NEGF) or single-time reduced density matrix methods. Approximations are introduced via a proper choice of the many-particle self-energy or decoupling of the BBGKY hierarchy. These approximations are based on Feynman's diagram approaches or on cluster expansions into single-particle and correlation operators.
Here, we develop a different approach where, instead of equations of motion for the many-particle NEGF (or density operators), single-time equations for the correlation functions of fluctuations are analyzed. We present a derivation of the first two equations of the alternative hierarchy of fluctuations and discuss possible decoupling approximations. In particular, we derive the polarization approximation (PA) which is shown to be equivalent to the single-time version [following by applying the generalized Kadanoff--Baym ansatz (GKBA)] of the nonequilibrium $GW$ approximation with exchange effects of NEGF theory, for weak coupling.
The main advantage of the quantum fluctuations approach is that the standard ensemble average can be replaced by a semiclassical average over different initial realizations, as was demonstrated before by Lacroix and co-workers [see e.g. D. Lacroix \textit{et al.}, Phys. Rev. B, 2014, \textbf{90}, 125112]. Here, we introduce the stochastic $GW$ (SGW) approximation and the stochastic polarization approximation (SPA) which are demonstrated to be equivalent to the single-time $GW$ approximation without and with exchange, respectively, in the weak coupling limit. In addition to the standard stochastic approach to sample initial configurations, we also present a deterministic approach. Our numerical tests confirm that our approach has the same favorable linear scaling with the computation time as the recently developed G1-G2 scheme [Schluenzen et al., Phys. Rev. Lett., 2020, \textbf{124}, 076601]. At the same time, the SPA and SGW approaches scale more favorably with the system size than the G1-G2 scheme, allowing to extend nonequilibrium $GW$ calculations to bigger systems.
%
%
\printkeywords
%
\end{abstract}

\section{Introduction}\label{s:intro}
The dynamics of quantum many-body systems following external excitation is of great interest in many areas such as dense plasmas, nuclear matter, ultracold atoms or correlated
solids. There is a large variety of methods available to simulate such systems which include real-time quantum Monte Carlo, density matrix renormalization group approaches, time-dependent density functional theory and quantum kinetic theory.

At present, only the formalisms of reduced density operators, e.g.,~\cite{bogolyubov_1946} and nonequilibrium Green's functions (NEGF), e.g.,~\cite{keldysh64, bonitz-etal.96jpcm, stefanucci-book,balzer-book} can rigorously describe the quantum dynamics of correlated systems in more than one dimension, e.g.,~\cite{schluenzen_prb16}. However, NEGF simulations are computationally expensive, among
other things, due to their cubic scaling with the simulation time $N_t$ (the number of time steps). Only recently,
linear scaling with $N_t$ could be achieved within the G1-G2 scheme~\cite{schluenzen_prl_20,joost_prb_20} which could be demonstrated even for advanced self-energies including the $GW$ and the $T$-matrix approximation. Even the nonequilibrium dynamically screened ladder approximation, which selfconsistently combines dynamical screening and strong coupling, is now feasible, at least for lattice models~\cite{joost_prb_22}. 

The advantage of time linear scaling of the G1-G2 scheme comes at a price: a simultaneous propagation of the single-particle and correlated two-particle Green's functions, $G_1$ and $\mathcal{G}_2$, requires a large computational effort for computing and storing all matrix elements of $\mathcal{G}_2$. For example, the CPU time of $GW$-G1-G2 simulations scales as $N_b^6$, where $N_b$ is the basis dimension. Even though this difficulty can be relieved using massively parallel computer hardware, it is well worth to look for alternative formulations of the problem that are more suitable for computations, ideally without loss of accuracy.

Here, we consider such an alternative formulation of the quantum many-body problem that is based on a stochastic approach to the (single-time) dynamics of quantum fluctuations. The idea is that correlation effects can be equivalently formulated in terms of correlation functions of fluctuations. This has been studied in detail, for classical systems by Klimontovich, see e.g.,~\cite{klimontovich_jetp_57, klimontovich_jetp_72, klimontovich_1982}, for an overview on his work see references~\cite{bonitz_cpp_03, dufty_jpcs_05}. For quantum systems, similar concepts were developed by Ayik, Lacroix and many others, e.g.,~\cite{ayik_plb_08, lacroix_prb14,lacroix_epj_14}. Before outlining our concept, we briefly mention how our approach differs from other stochastic methods that, of course, have a long history in physics. 

Fluctuations have been investigated in many areas of classics physics, including plasmas~\cite{rostoker61}, gases, liquids and the theory of liquid or plasma turbulence, e.g.,~\cite{kadomtsev68,sitenko-book}. Many-particle systems in thermodynamic equilibrium with interactions so strong that perturbation theories fail are successfully described using
classical and quantum Monte Carlo methods, e.g.,~\cite{foulkes_rmp_01,dornheim_physrep_18,filinov_ppcf_01}. Stochastic approaches have been also broadly applied to treat the time evolution of interacting classical systems, e.g., via molecular dynamics which are averaged over initial conditions,  kinetic equations including Monte Carlo collision techniques (PIC-MCC) or kinetic Monte Carlo methods. Similarly diverse is the spectrum of stochastic approaches to the dynamics of quantum systems which includes real-time quantum Monte Carlo,  various versions of quantum molecular dynamics, e.g.,~\cite{filinov96,filinov_prb_2}, path integral molecular dynamics, e.g.,~\cite{braams_jcp_06},
generalized Bohmian dynamics, e.g.,~\cite{Lardereaaw1634}, quantum corrections to semiclassical dynamics~\cite{polkovnikov_ap_10}, or the truncated Wigner approximation (TWA)~\cite{polkovnikov_pra03,Blakie_AdvancesInPhysics,Martin_2010} and its recently introduced improvements for spin models~\cite{pinna_prb_16, sundar_pra_19,kunimi_prr_21}. Such an extension of the TWA is presented, for example, in reference~\cite{Kastner_prb2016}, where the initial discrete Wigner distribution is sampled in phase space in combination with a cluster expansion for the phase space points. The realizations of the expansion coefficients are then propagated by a set of semiclassical equations based on the BBGKY hierarchy. A different approach for the description of correlation functions of fluctuations is given by Wetterich, e.g.,~\cite{Wetterich_prl1997}, where the time evolution of the generating functionals for the correlation functions is calculated from an exact functional differential equation instead of directly considering the correlation functions. For nonequilibrium situations, approximations in the form of an appropriate truncation of the generating functional are then applied in order to find solutions.

Here, we use an equation of motion approach that is similar to the single-time BBGKY hierarchy for reduced density operators, e.g.,~\cite{bogolyubov_1946}, and to the many-time Martin--Schwinger hierarchy of NEGF theory~\cite{martin_theory_1959}. However, instead of many-particle Green's functions or density operators, we consider an approach that is based on fluctuations of the single-particle Green's function operator on the time diagonal, $\delta \hat G(t)$, and products thereof, giving rise to single-time two-particle and higher fluctuations. We derive the first equations for the fluctuation hierarchy and compare them to the BBGKY hierarchy of the single-time reduced density operators to which they are fully equivalent. 
Further, we discuss possible closures to the hierarchy and concentrate on two  approximations that are equivalent to the single-time $GW$ approximation of many-particle theory with and without exchange which plays a fundamental role for ground state and nonequilibrium properties of quantum systems. 

While the fluctuation hierarchy has a similar complexity as the BBGKY hierarchy and the G1-G2 scheme, it has a major advantage: the correlated dynamics can be reformulated and mapped onto an effective single-particle problem, as was shown by Klimontovich, for classical systems~\cite{klimontovich_jetp_57}, and by Ayik and Lacroix, for quantum systems, within their stochastic mean field approximation~\cite{ayik_plb_08,lacroix_epj_14,lacroix_prb14}. 
The main result of our paper is the identification of two equations of motion for the  single-particle fluctuation $\delta \hat G$ --- the stochastic polarization approximation (SPA) and stochastic $GW$ (SGW) --- that are equivalent to single-time version of
$GW$ with and without exchange, respectively, in the weak coupling limit. By performing extensive tests for small Hubbard clusters, we verify this correspondence. Furthermore, we test the efficiency of this approach by utilizing various stochastic sampling schemes.

This paper is structured as follows. In section~\ref{s:theory} we briefly recall the formalism of second quantization and the equations of the G1-G2-scheme. In section~\ref{s:fluctuations} we introduce our concept of fluctuations of the nonequilibrium Green's functions and the resulting hierarchy of equations for the fluctuations. In section~\ref{s:approximations} we discuss the most important approximations to decouple the hierarchy. Section~\ref{s:SMF} is devoted to the stochastic mean field theory and section~\ref{s:sampling} to the different sampling approaches. Finally, section~\ref{s:application_to_Hubbard} is devoted to the application to the Hubbard model and contains our numerical results, and a summary is presented in section~\ref{s:discussion}.

\section{Theoretical framework}\label{s:theory}
Throughout this work we consider general nonequilibrium situations that are produced by an external excitation giving rise to complex time dependencies of all relevant quantities.
\subsection{Notation and definitions}
\label{ss:notation_and_definitions}
Although the general theory of NEGF is formulated in terms of functions depending on multiple time arguments, here we focus on the special case of equal times ($t=t'$) and use a simplified notation. For example, the single-particle correlation function will be denoted by $G^<_{ij} \equiv G^<_{ij}(t)\equiv G^<_{ij}(t,t)$, where we will suppress the time dependence.

In the following, the formalism of the second quantization is used, which is characterized by the bosonic/fermionic creation ($\hat{c}_i^\dagger$) and annihilation ($\hat{c}_i$) operators  and  a single-particle orbital basis $\{|1\rangle,\dots,|n\rangle\}$ of the underlying single-particle Hilbert space $\mathcal{H}$, which induces the so-called Fock space~$\mathcal{F}$ and allows for the treatment of states with a varying number of particles. Basic properties of the mentioned operators are the following \footnote{In this work, the upper/lower sign, i.e., ``$\pm$'' and ``$\mp$'', refers to bosons/fermions, respectively.}:
\begin{align}
    \Big[\hat{c}_i,\hat{c}_j^\dagger\Big]_{\mp}=\delta_{ij},\quad \Big[\hat{c}_i,\hat{c}_j\Big]_{\mp}=\Big[\hat{c}^\dagger_i,\hat{c}^\dagger_j\Big]_{\mp}=0. \label{eq:ladder_operators}
\end{align}
Here, a generic Hamiltonian is used to describe a quantum many-particle system, which is given by
\begin{equation}
    \hat{H}= \sum_{ij}h_{ij}\hat{c}^\dagger_i\hat{c}_j+\frac{1}{2}\sum_{ijkl}w_{ijkl}\hat{c}_i^\dagger\hat{c}^\dagger_j\hat{c}_l\hat{c}_k, \label{eq:generic_Hamiltonian}
\end{equation}
where the single-particle contribution is defined by $h_{ij}\coloneqq\langle i|\hat{T}+\hat{V}|j\rangle$. The operator $\hat{T}$ describes the kinetic energy and $\hat{V}$ an external potential that may be time dependent. For the second term we defined $w_{ijkl}\coloneqq\langle ij|\hat{W}|kl\rangle$, where $\hat{W}$ describes two-particle interactions. Note that $\hat{W}$ may also be time-dependent, e.g., when we compute a correlated initial state from an uncorrelated state, via the adiabatic switching procedure 
\cite{hermanns_prb14,joost_prb_22}. 

The starting point for NEGF theory is the one-body Green's function that depends on two time arguments situated on the Keldysh contour $\mathcal{C}$ which is defined as
\begin{equation}
    G_{ij}(z,z')=\frac{1}{\mathrm{i}\hbar}\left\langle \mathcal{T}_\mathcal{C}\left\{\hat{c}_i(z)\hat{c}^\dagger_j(z')\right\}\right\rangle,
\end{equation}
where $\mathcal{T}_\mathcal{C}$ is the time-ordering operator on the contour,  and averaging is performed with the correlated unperturbed density operator of the system. Details of the formalism can be found in textbooks, e.g.,~\cite{balzer-book,stefanucci-book} but will not be needed here. For us, it will be sufficient to consider the correlation functions $G^\gtrless$ for real time arguments on the time-diagonal $(t=t')$. The definition of these functions is again given by an ensemble average,
\begin{align}
    G^<_{ij}&=\pm\frac{1}{\mathrm{i}\hbar}\left\langle \hat{c}^\dagger_j\hat{c}_i\right\rangle, & G^>_{ij}&=\frac{1}{\mathrm{i}\hbar}\left\langle \hat{c}_i\hat{c}^\dagger_j\right\rangle, \label{eq:g</>_def}\\
    \hat{G}^<_{ij}&=\pm\frac{1}{\mathrm{i}\hbar} \hat{c}^\dagger_j\hat{c}_i, & \hat{G}^>_{ij}&=\frac{1}{\mathrm{i}\hbar} \hat{c}_i\hat{c}^\dagger_j \label{eq:g</>_op_def}.
\end{align}
On the real-time-diagonal, the lesser component is, therefore, proportional to the single-particle density matrix, $n_{ij}\coloneqq \langle\hat{c}^\dagger_j\hat{c}_i\rangle = \pm \mathrm{i}\hbar G^<_{ij}$. 
Note that, in the second line, equation~(\ref{eq:g</>_op_def}), we omit the ensemble averaging. The corresponding Green's function operators will be denoted ${\hat G}^\gtrless$ and are the starting point of our theory. Here, we will not consider bosons in the condensate, but an extension of the approach is straightforward and would involve also anomalous correlators.

Analogously to the single-particle NEGF, higher order Green's functions can be defined, but since only the limit of equal times is of interest in this work, only one special case of the two-particle Green's function on the time-diagonal will be considered here,
\begin{equation}
    G^{(2),<}_{ijkl}\coloneqq -\frac{1}{\hbar^2}\left\langle \hat{c}^\dagger_k\hat{c}^\dagger_l\hat{c}_j\hat{c}_i\right\rangle.
\end{equation}
This quantity deviates from the standard two-particle density matrix by an additional prefactor of $(-1/\hbar^2)$. Furthermore, we consider a decomposition of the two-particle Green's function on the time-diagonal into a mean-field (Hartree--Fock) and a correlation contribution,
\begin{equation}
    G^{(2),<}_{ijkl}=G^<_{ik}G^<_{jl}\pm G^<_{il}G^<_{jk}+\mathcal{G}_{ijkl} \label{eq:G2_cluster},
\end{equation}
where the first two terms are the Hartree and Fock contributions, whereas $\mathcal{G}$ describes two-particle correlations. For reduced density matrices, this decomposition is called cluster expansion~\cite{bonitz_qkt}. The correlation contribution is the central quantity of the so-called G1-G2 scheme, as introduced in~\cite{schluenzen_prl_20, joost_prb_20}, and obeys the following (pair-)exchange symmetries,
\begin{align}
    \mathcal{G}_{ijkl}&=\mathcal{G}_{jilk}=\Big[\mathcal{G}_{klij}\Big]^*=\pm\mathcal{G}_{jikl}=\pm\mathcal{G}_{ijlk} \label{eq:G2_symmetries}.
\end{align}
\subsection{Time-diagonal Keldysh--Kadanoff--Baym equation} \label{ss:tdKBE}
The equations of motion (EOMs) of the NEGF are the Keldysh--Kadanoff--Baym equations (KBE), which contain the single-particle as well as the two-particle NEGF, where the latter is usually approximated by various self-energy approximations~\cite{stefanucci-book,balzer-book,schluenzen_jpcm_19}. As was shown in references~\cite{schluenzen_prl_20,joost_prb_20}, the EOM for the lesser component of the NEGF on the time-diagonal is given by the time-diagonal KBE and can be written as \footnote{For two single-particle quantities $A$ and $B$, we define the the commutator as $[A,B]_{ij}\coloneqq \sum_k\{ A_{ik}B_{kj}-B_{ik}A_{kj}\}$.} 
\begin{equation}
    \mathrm{i}\hbar\partial_tG^<_{ij}=\Big[h^\mathrm{HF},G^<\Big]_{ij}+\left[{I}^{(\mathcal{G})}+{I}^{(\mathcal{G})\dagger}\right]_{ij}, \label{eq:EOM_G1_corr}
\end{equation}
where the Hartree--Fock contribution of the two-particle Green's function is included in an effective single-particle Hartree--Fock Hamiltonian, which is defined as
\begin{equation}
    h^\mathrm{HF}_{ij}\coloneqq h_{ij}\pm\mathrm{i}\hbar\sum_{kl}w^\pm_{ikjl}G_{lk}^<, \label{eq:hartree-fock-hamiltonian}
\end{equation}
where we introduced an (anti-)symmetrized interaction tensor
\begin{equation}
    w^\pm_{ijkl}\coloneqq w_{ijkl}\pm w_{ijlk}. \label{eq:antisymm_potential}
\end{equation}
The last term in equation~\eqref{eq:EOM_G1_corr} corresponds to the collision integral and couples to the correlation part of the two-particle Green's function. It is given by
\begin{equation}
    {I}^{(\mathcal{G})}_{ij}\coloneqq \pm\mathrm{i}\hbar\sum_{klp}w_{iklp}\mathcal{G}_{lpjk}.\label{eq:collision_term_corr}
\end{equation}
Equation~\eqref{eq:EOM_G1_corr} is the first equation of the G1-G2 scheme and requires the solution of an EOM for $\mathcal{G}$, which can be derived from the known self-energy approximations and applying the GKBA with Hartree--Fock propagators~\cite{joost_prb_20,joost_prb_22}. An alternative derivation can be given starting from the single-time BBGKY hierarchy~\cite{bonitz_qkt}. A detailed comparison of both apporaches was given in reference~\cite{joost_prb_22}. Here, we reproduce the system of single-time equations that was derived in reference~\cite{joost_prb_22} and explain the terms in the equations as well as the approximations applied in the present paper,
%
\begin{align}
\mathrm{i}\hbar \partial_t 
\mathcal{G}_{ijkl} -& [h^{(2),\rm HF}, \mathcal{G}]_{ijkl}
= \Psi^\pm_{ijkl} + L_{ijkl}+ P^\pm_{ijkl}+ C^{(3)}_{ijkl},
\label{eq:g2-equation}
\\
 h^{(2),\rm HF}_{ijkl} =&h_{ik}^{\rm HF}\delta_{jl} + h_{il}^{\rm HF}\delta_{ik},\label{eq:g1g2-h2hf}\\
\Psi^\pm_{ijkl}  \coloneqq& (\mathrm{i}\hbar)^2\sum_{pqrs}w^\pm_{pqrs}\Big\{G^>_{ip}G^>_{jq}G^<_{rk}G^<_{sl}-(>\leftrightarrow <)\Big\},\label{eq:source_term_corr}
\\
L_{ijkl} =& \sum_{rs}\left\{
{V}_{ijrs}\,\mathcal{G}_{rskl} -  \mathcal{G}_{ijrs}\,{V}^*_{rskl}\right\},
\label{eq:l-def}
\\
P^\pm_{ijkl} =& \Pi^\pm_{ijkl} \pm \Pi^\pm_{ijlk},
\label{eq:ppm-def}\\
     \Pi^\pm_{ijkl} \coloneqq& \pm\mathrm{i}\hbar \sum_{pqr}\left\{\mathcal{G}_{rjpl}(w^\pm_{ipqr}G^<_{qk}-w^\pm_{qpkr}G^<_{iq}) + \mathcal{G}_{iqkp}(w^\pm_{pjqr}G^<_{rl}-w^\pm_{prql}G^<_{jr})\right\},\label{eq:pi-def}\\
   C^{(3)}_{ijkl}\coloneqq&\pm\mathrm{i}\hbar\sum_{pqr}\left\{w_{ipqr}\mathcal{G}^{(3)}_{rqjpkl}+w_{pjqr}\mathcal{G}^{(3)}_{iqrkpl} - w_{pqkr}\mathcal{G}^{(3)}_{irjpql}-w_{pqrl}\mathcal{G}^{(3)}_{ijrkqp}\right\}, \label{eq:def-c3}
\\ 
V_{ijkl} =& \sum_{rs}(\delta_{ir}\delta_{jl}\pm n_{ir}\delta_{js} \pm n_{js}\delta_{ir})w_{rskl}.
\end{align}
Here, $h^{(2),\rm HF}$, equation~\eqref{eq:g1g2-h2hf}, denotes the two-particle Hartree--Fock Hamiltonian, $\Psi^\pm$ is an inhomogeneity (it does not depend on $\mathcal{G}$) that gives rise to direct and exchange second order Born scattering terms. The term $L$, equation~\eqref{eq:l-def}, contains the ladder diagrams that are relevant for strong coupling effects (particle-particle $T$-matrix approximation, TPP), whereas $\Pi^{\pm}$, equation~\eqref{eq:pi-def}, comprises all polarization diagrams that describe the dynamical screening effects. Note that the term $P^\pm_{ijkl}$, equation~\eqref{eq:ppm-def}, in addition to $\Pi^\pm_{ijkl}$, contains a term $\Pi^\pm_{ijlk}$ which represents particle-hole $T$-matrix  (TPH) diagrams~\cite{joost_prb_22}. Finally, $C^{(3)}$ couples to the rest of the system via three-particle correlations, $\mathcal{G}^{(3)}$. 

In what follows, a central role is played by the $GW$ approximation of many-particle physics. Its single-time version is recovered in the G1-G2-scheme, equation~\eqref{eq:g2-equation}, if the three--particle correlations, $C^{(3)}$, the ladder term $L$ and the particle-hole $T$-matrix contribution, $\Pi^\pm_{ijlk}$, are neglected. The remaining polarization term $\Pi^\pm_{ijkl}$ then corresponds to $GW$ with exchange corrections which will be denoted $GW^\pm$. On the other hand, standard $GW$ (without exchange diagrams) follows by replacing $\Psi^\pm \to \Psi$ and $\Pi^\pm_{ijkl}\to \Pi$, which means that, in these terms, the replacement $w^\pm \to w$ is made~\cite{bonitz_qkt, joost_prb_20, joost_prb_22}. In the present paper we present stochastic versions of $GW$ and $GW^\pm$.

\section{Quantum fluctuations of the Green's functions} \label{s:fluctuations}
Instead of considering the correlations via the correlated part of the two-particle Green's function,~$\mathcal{G}$, we now take a different approach that is analogous to the classical fluctuation approach that was developed by Klimontovich~\cite{klimontovich_1982}. He considered fluctuations of the microscopic phase space density~\cite{klimontovich_1982}, $\delta N(\textbf{r},\textbf{p},t) \equiv N(\textbf{r},\textbf{p},t) -n f(\textbf{r},\textbf{p},t)$, where $n$ is the density and $f$ is the single-particle probability density in phase space that obeys a standard kinetic equation. Here, we generalize this approach to quantum many-particle systems, fully including quantum effects and spin statistics. 

The natural starting point of our quantum fluctuation approach is then to replace $n f \to G^<$ and $N \to \hat G^<$. Consequently, the defining quantity is the single-particle fluctuation operator,
\begin{equation}
    \delta\hat{G}_{ij}\coloneqq\hat{G}^<_{ij}-G^<_{ij}, \label{eq:deltag-def_1}
\end{equation}
where, obviously, $G^<_{ij}=\langle \hat{G}^<_{ij} \rangle$ and $\langle \delta\hat{G}_{ij} \rangle =0$.
The attractive feature of this approach is that 
 two- and three-particle fluctuation operators emerge simply as products of single-particle fluctuation operators. In what follows, we consider in detail the two-particle and three-particle fluctuations which are expectation values of the operator products,
 \begin{align}
     \gamma_{ijkl}&\coloneqq \left\langle \fluc_{ik}\fluc_{jl}\right\rangle, \label{eq:2p_fluc_def}\\
     \Gamma_{ijklmn} &\coloneqq \left\langle \fluc_{il}\fluc_{jm}\fluc_{kn}\right\rangle.  \label{eq:3p_fluc_def}
 \end{align}
 
 Before proceeding with an analysis of the properties and equations of motion of these quantities, we note that analogous quantities 
were analyzed in the theory of reduced density matrices where they are called ``correlation matrices''~\cite{Valdemoro2009}. 
Therein, an operator $\hat{\mathcal{Q}}$ is defined as the projector on the subspace orthogonal to the $N$-particle state $|\Psi\rangle$, 
\begin{equation}
    \hat{\mathcal{Q}}\coloneqq\sum_{\Psi'\neq \Psi}|\Psi'\rangle\langle \Psi'|\coloneqq\hat{\mathrm{I}}-|\Psi\rangle\langle \Psi|,
\end{equation}
where $\hat{\mathrm{I}}$ denotes the identity operator. 
It is straightforward to express our two-particle fluctuations in terms of the projector $\hat{\mathcal{Q}}$:
\begin{equation}
    \gamma_{ijkl}=\left\langle \hat{G}_{ik}\hat{\mathcal{Q}}\hat{G}_{jl}\right\rangle. \label{eq:fluc_alt_def}
\end{equation}
This alternative expression shows that fluctuations describe correlation effects of $N$ particles which undergo virtual excitations and de-excitations, in order to avoid each other. 
The main drawback of this approach is, however, that it is incapable of describing single-particle fluctuations, as defined in equation~\eqref{eq:deltag-def_1}. Therefore, we proceed with our approach that is based on the fluctuation operator~\eqref{eq:deltag-def_1}. 

\subsection{Properties of the single-particle and two-particle fluctuations} \label{ss:properties}
Since only the limit of equal times is considered here, there exists only a single one-particle fluctuation operator [cf. equations \eqref{eq:ladder_operators}, \eqref{eq:g</>_def} and \eqref{eq:g</>_op_def}], 
\begin{align}
\delta \hat{G}_{ij} = \hat{G}^>_{ij}-G^>_{ij}=\hat{G}^<_{ij}-G^<_{ij}.
\end{align}
Thus, fluctuations of the greater and lesser components of the single-particle Green's function coincide on the time-diagonal. This means that, in the case of fermions, the sum of fluctuations of particles and holes vanishes. 

Using the properties of the creation and annihilation operators defined in equation~\eqref{eq:ladder_operators}, the  two-particle Green's function can be expressed in terms of two-particle fluctuations~\cite{Valdemoro2009},
\begin{equation}
    G^{(2),<}_{ijkl}=G^<_{ik}G^<_{jl} \mp \frac{1}{\mathrm{i}\hbar}\delta_{il}G^<_{jk}+\gamma_{ijkl}. \label{eq:G^2,<_fluc}
\end{equation}
Now, using relation~\eqref{eq:G2_cluster}, we establish the relation between two-particle correlations and fluctuations:
\begin{align}
    \gamma_{ijkl} &= \mathcal{G}_{ijkl} \pm G^>_{il}G^<_{jk}\nonumber \\
    &=\mathcal{G}_{ijkl} \pm \gamma^s_{ijkl}.
    \label{eq:G2_fluc}
    \end{align}
This shows what the Kronecker delta term in equation~\eqref{eq:G^2,<_fluc} physically means: it contains the exchange contribution and additionally a term describing particle-hole polarization effects~\cite{Valdemoro2009}. Analogously to the classical case~\cite{klimontovich_1982}, we define the difference of correlations and fluctuations as ``source term'',
\begin{equation}
    \gamma_{ijkl}^s\coloneqq  G^>_{il}G^<_{jk}.
    \label{eq:def-gamma-s}
\end{equation}
This source term can be interpreted as simultaneous excitation of a particle-hole pair: a particle is annihilated in orbital $k$ and created in orbital $j$ whereas a hole is annihilated in orbital $i$ and created in~$l$. Such ``particle-hole fluctuations'' can be understood as the basis for all fluctuations. They are always present in a quantum system, even if the particles are non-interacting \footnote{They arise from vacuum fluctuations. In a relativistic system, the same source term would describe particle-antiparticle pair excitations.}. In the latter case, $\mathcal{G}=0$ and the total fluctuation $\gamma$ coincides with $\gamma^s$. Of course, in the presence of correlations, the latter will affect the dynamics of the source term. The standard example for e-h pairs is the electron gas at low temperature. Therein, spontaneous interaction with a photon of energy $\omega$ and wave number $q$ may excite an electron from a state $p$ (below the Fermi energy) to a state $p+q$ above it. The probability of this process is proportional to $n_p(1-n_{p+q})$, ($n_p$ is the momentum distribution), in agreement with the definition~\eqref{eq:def-gamma-s}. 

By construction, the two-particle correlation function $\mathcal{G}$ obeys the exchange symmetries~\eqref{eq:G2_symmetries}. However, fluctuations in general do not obey these symmetries. 
By using the relation between correlations and fluctuations, equation~\eqref{eq:G2_fluc}, it is straighforward to establish the symmetries of the 
two-particle fluctuations under the exchange of particles\footnote{Note that, in the classical case, the fluctuations obey the exchange symmetry.},
\begin{align}
    \gamma_{ijkl}\mp\gamma_{ijkl}^s&=\gamma_{jilk}\mp \gamma^s_{jilk}=\pm \gamma_{jikl}-\gamma^s_{jikl}=\pm \gamma_{ijlk}-\gamma^s_{ijlk}. \label{eq:gamma-gamma}
\end{align}
Lastly, we find that two-particle fluctuations obey the following exchange symmetry,
\begin{equation}
    \gamma_{ijkl}=\Big[\gamma_{lkji}\Big]^*.
\end{equation}

\subsection{Single-particle dynamics} \label{ss:dynamics_1p_fluc}
Using equation~\eqref{eq:G2_fluc}, it is straightforward to express the time-diagonal KBE [cf. equation~\eqref{eq:EOM_G1_corr}] in terms of two-particle fluctuations,  
\begin{equation}
    \mathrm{i}\hbar\partial_tG^<_{ij}=\Big[h^{\mathrm{H}},G^<\Big]_{ij}+\Big[I^{(\gamma)}+I^{(\gamma)\dagger}\Big]_{ij}\label{eq:EOM_G1_fluc}, 
\end{equation}
where we introduced an effective single-particle Hartree Hamiltonian, 
\begin{equation}
     h^{\mathrm{H}}_{ij}\coloneqq h_{ij}\pm\mathrm{i}\hbar\sum_{kl}w_{ikjl}G^<_{lk}, \label{eq:hartree-hamiltonian}
\end{equation}
and a collision term,
\begin{equation}
    I^{(\gamma)}_{ij}\coloneqq\pm\mathrm{i}\hbar\sum_{klp} w_{iklp}\gamma_{plkj}.\label{eq:collision_term_fluc}
\end{equation}
Note that equations~\eqref{eq:EOM_G1_corr} and~\eqref{eq:EOM_G1_fluc} differ with respect to the effective single-particle Hamiltonians, as defined in equations~\eqref{eq:hartree-fock-hamiltonian} and~\eqref{eq:hartree-hamiltonian}. This is a feature of fluctuations, which, unlike Green's functions or density matrices, do not obey symmetries under exchange of particles. Likewise, effective interactions of fluctuations do not contain any exchange contributions (no Fock term). As a consequence, quantum fluctuations behave similar to classical fluctuations~\cite{klimontovich_1982}.
Finally, the collision terms given by equations~\eqref{eq:collision_term_corr} and~\eqref{eq:collision_term_fluc}, which lead to the coupling to higher order contributions, have the same structure but involve  different two-particle quantities, i.e., $\mathcal{G}$ and $\gamma$, respectively.
 
To close equation~\eqref{eq:EOM_G1_fluc} we now need an equation for $\gamma$. To derive this equation, we start with deriving an EOM of the single-particle fluctuation operator. This can be done by subtracting the EOMs for $\hat{G}^<$ and $G^<$ from each other, where the EOM for $\hat{G}^<$ is given by
\begin{align}
    &\mathrm{i}\hbar\partial_t\hat{G}^<_{ij} =\left[\hat{h}^\mathrm{H}, \hat{G}^<\right]_{ij}, \label{eq:EOM_G_op}\\
    &\hat{h}^\mathrm{H}_{ij}\coloneqq h_{ij}\pm\mathrm{i}\hbar\sum_{kl}w_{ikjl}\hat{G}^<_{lk}. \label{eq:hartree-hamiltonian_op}
\end{align}
Subtracting equations~\eqref{eq:EOM_G_op} and~\eqref{eq:EOM_G1_fluc} and replacing all operators by their expectation values and fluctuations, as in equation~\eqref{eq:deltag-def_1}, leads to the following EOM for the single-particle fluctuation operator:
\begin{align}
    \mathrm{i}\hbar\partial_t\delta\hat{G}_{ij}=&\mathrm{i}\hbar\partial_t\left\{\hat{G}^<_{ij}-G^<_{ij}\right\}= \left[h^\mathrm{H},\delta\hat{G}\right]_{ij}+\left[\delta\hat{U}^\mathrm{H},G^<\right]_{ij}+\left[\delta\hat{U}^\mathrm{H},\delta\hat{G}\right]_{ij}- \Big[I^{(\gamma)}+I^{(\gamma)\dagger}\Big]_{ij},\label{eq:EOM_deltaG}
\end{align}
where we defined the operator of an effective single-particle Hartree potential induced by fluctuations,
\begin{equation}
    \delta\hat{U}^\mathrm{H}_{ij}\coloneqq\pm\mathrm{i}\hbar\sum_{kl}w_{ikjl}\delta\hat{G}^<_{lk}. \label{eq:fluc-hartree-potential}
\end{equation}
Therefore, the corresponding commutator represents the interaction of $G^<$ and a ``fluctuation mean-field''. 
Furthermore, the last two terms in equation \eqref{eq:EOM_deltaG} correspond to fluctuations of fluctuations (second-order fluctuations), which can be seen by identically rewriting the collision term, equation~\eqref{eq:collision_term_fluc}, as
\begin{equation}
    \Big[I^{(\gamma)}+I^{(\gamma)\dagger}\Big]_{ij}=\left\langle\left[\delta\hat{U}^\mathrm{H},\delta\hat{G}\right]\right\rangle_{ij}. 
\end{equation}
Thus, the last two terms in equation~\eqref{eq:EOM_deltaG} can be expressed as
\begin{align}
    \delta\left(\left[\delta\hat{U}^\mathrm{H},\delta\hat{G}\right]_{ij}\right)=\pm\mathrm{i}\hbar\sum_{klp} \left\{w_{iklp}\delta\hat{\gamma}_{plkj}-w_{kljp}\delta\hat{\gamma}_{ipkl}\right\}. \label{eq:2nd_order_fluc}
\end{align}
While expressing these terms as second-order fluctuations is helpful for physical interpretation, this form turns out to be mathematically not advantageous.
Instead of an EOM for $\delta \gamma$, we consider equations for $\gamma$ and $\delta\hat G$. Remarkably, by using just the EOM of $\delta\hat{G}$ [cf. equation~\eqref{eq:EOM_deltaG}], it is possible to derive the EOM of any $N$-particle fluctuation, which makes this equation the cornerstone of our formalism.

\subsection{Dynamics of two-particle fluctuations $\gamma$} \label{ss:dynamics_2p_fluc}

To close equation \eqref{eq:EOM_G1_fluc}, we need an EOM for $\gamma$. It follows directly from equation~\eqref{eq:EOM_deltaG} which is multiplyied by another factor $\delta \hat G$ and by a subsequent ensemble average\footnote{For two two-particle quantities $A$ and $B$, we define the commutator as $[A,B]_{ijkl}\coloneqq\sum_{pq}\left\{ A_{ijpq} B_{pqkl}-B_{ijpq}A_{pqkl}\right\}$.}:
\begin{align}
    \mathrm{i}\hbar\partial_t\gamma_{ijkl} &=\left[ h^{(2),\rm{H}},\gamma\right]_{ijkl}+\pi_{ijkl}+C^{(\Gamma)}_{ijkl}, \label{eq:EOMgamma}\\
    h^{(2),\mathrm{H}}_{ijkl} &\coloneqq h^\mathrm{H}_{ik}\delta_{jl}+h^\mathrm{H}_{jl}\delta_{ik},
\end{align}
where we introduced an effective two-particle Hartree Hamiltonian.
The second term on the r.h.s. in equation~\eqref{eq:EOMgamma}, is a polarization contribution involving fluctuations,
\begin{align}
     \pi_{ijkl} \coloneqq &\pm\mathrm{i}\hbar \sum_{pqr}\gamma_{rjpl}\left\{w_{ipqr}G^<_{qk}-w_{qpkr}G^<_{iq}\right\}\nonumber\\
&\pm\mathrm{i}\hbar\sum_{pqr}\gamma_{iqkp}\left\{w_{pjqr}G^<_{rl}-w_{prql}G^<_{jr}\right\}, \label{eq:2p_polarization_contribution}
\end{align}
which is similar to the polarization term $\Pi^\pm$ of the G1-G2-scheme, cf. equation~\eqref{eq:ppm-def}. However, in contrast to the latter, equation~\eqref{eq:2p_polarization_contribution} does not contain any exchange contributions (it contains the pair potential $w$ instead of $w^\pm$) and also misses the particle-hole ladder term.
Using the property~\eqref{eq:gamma-gamma}, it is possible to rewrite $\pi$ such that   $G^<$ is eliminated:
\begin{align}
    \pi_{ijkl}=&(\mathrm{i}\hbar)^2\sum_{pqrs}w_{pqrs}\left\{\gamma_{iskq}\gamma_{jrlp}-\gamma_{rjpl}\gamma_{siqk}\right\}. \label{eq:2p_polarization_contribution_2}
\end{align}
Finally, the last term on the r.h.s. in equation \eqref{eq:EOMgamma}, couples to three-particle fluctuations,
\begin{align}
   C^{(\Gamma)}_{ijkl}\coloneqq&\pm\mathrm{i}\hbar\sum_{pqr}\left\{w_{ipqr}\Gamma_{rqjpkl}+w_{pjqr}\Gamma_{iqrkpl}\right\} \nonumber \\
    &\mp\mathrm{i}\hbar\sum_{pqr}\left\{w_{pqkr}\Gamma_{irjpql}+w_{pqrl}\Gamma_{ijrkqp}\right\}, \label{eq:def-cgamma}
\end{align}
which is again similar to the corresponding term $C^{(3)}$, equation~\eqref{eq:def-c3} in~equation~\eqref{eq:g2-equation} that involves the three-particle correlation function. Thus, similar to the equations for the reduced density operators which form the BBGKY hierarchy, we again obtain a hierarchy of equations for the fluctuations. We underline that the system of equations~\eqref{eq:EOM_G1_fluc} and~\eqref{eq:EOMgamma} is still exact provided that the three-particle fluctuation~$\Gamma$ is known. Interestingly, equation~\eqref{eq:EOMgamma} does not explicitly contain exchange, second-order Born and $T$-matrix contributions giving rise to a simpler formal structure. Those missing terms are, of course, ``hidden'' in the three-particle fluctuations $\Gamma$. The relation between $\gamma$ and $\mathcal{G}$ is further illustrated in section \ref{s:approximations} and requires further investigation of the three-particle contributions. 
\section{Approximations} \label{s:approximations}
Similar to the case of the single-particle BBGKY hierarchy, we  now explore possible approximations to decouple the fluctuation hierarchy. We also analyze how the  approximations known from the BBGKY formalism, e.g.,~\cite{joost_prb_22}, can be recovered in the present scheme and what new approximations might be offered by the fluctuations approach. For the latter, it is interesting to compare to the classical case.
\subsection{Approximation of first moments} \label{ss:approximation_of_first_moments}
The simplest approximations that were introduced in the classical case are the so-called ``approximations of moments''~\cite{klimontovich_1982}, where either two- or three-particle fluctuations are neglected. In the case $\gamma\rightarrow 0$, i.e., the approximation of first moments, we find a closed equation for $G^<$, which is a mean-field (Hartree or nonlinear Vlasov) equation,
\begin{equation}
    \mathrm{i}\hbar\partial_t G_{ij}^<=\Big[h^\hh,G^<\Big]_{ij}. \label{eq:EOM_G1_vlasov}
\end{equation}
Since equation~\eqref{eq:EOM_G1_vlasov} contains no Fock term, this approximation is only reasonable if the exchange effects are of minor importance such as for weakly degenerate systems.
\subsection{Time-dependent Hartree--Fock approximation} \label{ss:TDHF}
The missing Fock term in equation~\eqref{eq:EOM_G1_vlasov} is restored if the collision integral is not neglected, as in the approximation of first moments, but treated in the following approximation: we assume that two-particle correlations, $\mathcal{G}$, are negligible, compared to the source contributions, i.e., $\gamma_{ijkl}\approx \pm \gamma_{ijkl}^s$. Inserting this expression into equation~\eqref{eq:collision_term_fluc} leads to the well-known Hartree--Fock approximation (TDHF),
\begin{equation}
    \mathrm{i}\hbar\partial_t G^<_{ij}=\Big[h^\mathrm{HF},G^<\Big]_{ij}.
\end{equation}
This further shows that fluctuations contain a variety of additional effects that are not related to correlations and, thus, fluctuations are generally not negligible, even in the case of weak coupling. Further note that this is different from the classical case. Therein, the source term $\gamma^s$ does not contribute to the collision integral at all. This difference is not surprising because exchange (Fock term) is a pure quantum effect.
\subsection{Approximation of second moments (2M)} \label{ss:approximation_of_second_moments}
The approximation of second moments follows with setting $\Gamma\rightarrow 0$, with the result
\begin{align}
    \mathrm{i}\hbar\partial_t\gamma^\mathrm{2M}_{ijkl}=\left[ h^{(2),\rm{H}},\gamma^\mathrm{2M}\right]_{ijkl}+\pi_{ijkl} \label{eq:EOM_second_moment_approximation}.
\end{align}
In order to investigate this approximation further, we need to reconsider the relation between two-particle correlations and fluctuations and find, for the polarization term [cf. equation~\eqref{eq:2p_polarization_contribution}],
\begin{equation}
    \pi_{ijkl}=\Pi_{ijkl}+\Psi_{ijkl}. \label{eq:fluc_polarization_comp}
\end{equation}
Here, $\Pi$ denotes the polarization contribution \eqref{eq:pi-def}, whereas $\Psi$ is the inhomogeneity, equation~\eqref{eq:source_term_corr}, in the EOM of $\mathcal{G}$. Note that the superscript ``$\pm$'' is missing here which indicates that no (anti-)symmetrized potential appears, i.e., $w^\pm \to w$. While still accounting for Pauli blocking, exchange effects are not fully incorporated, similar to the case of the approximation of first moments. 

The appearance of the $\Psi$ and $\Pi$ terms in equation~\eqref{eq:EOM_second_moment_approximation} indicates that the 2M-approximation  already contains parts of the standard $GW$ contributions, as given within the G1-G2 scheme (for the correspondence to the $GW$ approximation, see the end of section~\ref{ss:tdKBE}). To identify the missing contributions, we consider the source term and its dynamics which are derived from equation~\eqref{eq:EOM_G1_fluc} by using $\partial_t G^<=\partial_t G^>$,
\begin{align}
    &\mathrm{i}\hbar\partial_t\gamma_{ijkl}^{{\rm 2M},s} =\left[ h^{(2),\rm{H}},\gamma^{{\rm 2M}, s}\right]_{ijkl}+ R^{(\gamma)}_{ijkl},\label{eq:EOM_gamma^s_fluc}\\
    &R^{(\gamma)}_{ijkl} \coloneqq   G^>_{il}\Big[  I^{(\gamma)} +  I^{(\gamma)\dagger}\Big]_{jk}+\Big[  I^{(\gamma)} +  I^{(\gamma)\dagger}\Big]_{il}G^<_{jk}, \quad
    \label{eq:residual_term} 
\end{align}
where $R^{(\gamma)}$ is a residual term.
Alternatively, equation~\eqref{eq:EOM_gamma^s_fluc} can be rewritten with the Hartree--Fock Hamiltonian in the commutator by using two-particle correlations instead of fluctuations, taking into account equation~\eqref{eq:EOM_G1_corr}: 
\begin{equation}
    \mathrm{i}\hbar\partial_t \gamma_{ijkl}^{{\rm 2M},s}=\left[ h^{(2),\rm{HF}},\gamma^{{\rm 2M},s}\right]_{ijkl}+{R}^{(\mathcal{G})}_{ijkl}, \label{eq:EOM_gamma^s_corr}
\end{equation}
where the residual term $R^{(\mathcal{G})}$ follows from $R^{(\gamma)}$ by replacing $\gamma \to \mathcal{G}$, in the collision integrals.
It is important to note that this equation accounts for the otherwise missing Fock term.

The final step for the comparison of the 2M approximation, equation~\eqref{eq:EOM_second_moment_approximation}, with the $GW$ approximation, is to add the equations for $\gamma^{\rm 2M}$ and $\gamma^{{\rm 2M},s}$, taking into account relation~\eqref{eq:G2_fluc} and denoting $\mathcal{G}^{\rm 2M}=\gamma^{\rm 2M}\mp\gamma^{{\rm 2M},s}$, 
\begin{align}
    \mathrm{i}\hbar\partial_t\mathcal{G}^\mathrm{2M}_{ijkl}=&\left[h^{(2),\mathrm{H}},\mathcal{G}^\mathrm{2M}\right]_{ijkl}+\Psi_{ijkl}+\Pi_{ijkl}\mp {R}^{(\mathcal{G})}_{ijkl}\mp \left[U^{(2),\mathrm{F}},\gamma^{{\rm 2M},s}\right]_{ijkl},
\end{align}
where we defined $U^{(2),\mathrm{F}}\coloneqq h^{(2),\mathrm{HF}}-h^{(2),\mathrm{H}}$.
We conclude that the EOM \eqref{eq:EOM_second_moment_approximation} for $\gamma^\mathrm{2M}$ is missing exchange contributions to the effective two-particle Hartree--Fock Hamiltonian and terms that compensate for ${R}^{(\mathcal{G})}$ in order to be fully equivalent to the $GW$ approximation. 

Whether the 2M approximation has an area of application by itself is not known at the moment and requires further analysis. However, in what follows we concentrate  on the $GW$ approximation. In the next paragraph, we identify the  terms that are missing so far, to reproduce the $GW$ and the $GW^\pm$
 approximations.

\subsection{Quantum polarization approximation} \label{ss:quantum_polarization_approximation}
 Next, we consider the quantum analogue of the classical polarization approximation (PA) which is known to be equivalent to the Balescu--Lenard equation ~\cite{klimontovich_1982}. This, in turn, is the classical limit of the nonequilibrium $GW$ approximation~\cite{bonitz_qkt}. 
\subsubsection{Approximation for the three-particle fluctuations} \label{sss:polarization_approximation_2p}
In the classical case, the PA is derived by assuming
that three-particle correlations are negligible, 
and  two-particle correlations are small compared to a product of two single-particle distributions.  
This motivates us to use the following three weak coupling approximations:
\begin{description}
\item[i)] $|\mathcal{G}^{(3)}|\rightarrow0$,
\item[ii)] $|\mathcal{G}|\ll|\gamma|$,
\item[iii)] $|\mathcal{G}|\ll|\gamma^s|$.
\end{description}
The task is now to identify the relevant approximation for the three-particle term $C^{(\Gamma)}$ in the equation for~$\gamma$, since this term is the source of the terms we are missing so far.

We start by considering the equation of motion of the two-particle source term, equation~\eqref{eq:EOM_gamma^s_corr}. Applying condition iii) leads to a neglect of the residual term $R^{(\mathcal{G})}$,

\begin{equation}
    \mathrm{i}\hbar\partial_t \gamma_{ijkl}^{\mathrm{PA},s}=\left[ h^{(2),\rm{HF}},\gamma^{\mathrm{PA},s}\right]_{ijkl}. \label{eq:EOM_gamma^s_P}
\end{equation}
Next, we return to the full equation for the two-particle fluctuation, equation~\eqref{eq:EOMgamma}, and establish a relation between three-particle correlations, $\mathcal{G}^{(3)}$, and fluctuations, $\Gamma$, which is analogous to equation~\eqref{eq:G2_fluc}:
\begin{align}
    \mathcal{G}^{(3)}_{ijklmn} &=\Gamma_{ijklmn}-G^>_{im}G^>_{jn}G^<_{kl}-G^>_{in}G^<_{jl}G^<_{km}\nonumber\\ 
    &\mp G^<_{jl}\mathcal{G}_{ikmn} \mp G^<_{km}\mathcal{G}_{ijln}- G^<_{kl}\mathcal{G}_{ijmn}\nonumber\\ 
    &\mp G^>_{im}\mathcal{G}_{jkln}\mp G^>_{jn}\mathcal{G}_{iklm}-G^>_{in}\mathcal{G}_{jklm}, \label{eq:G3_Gamma_P}
\end{align}
where the detailed derivation is given in Appendix~\ref{app:3p_corr_fluc_relation_derivation}.

Equation \eqref{eq:G3_Gamma_P} is now simplified taking into account approximations i), ii) and iii). Here, it is important to note that the single-particle Green's functions in the first line of equation~\eqref{eq:G3_Gamma_P} allow for different source terms to be considered, e.g., in the first term, $G^>_{im}G^>_{jn}G^<_{kl}=G^>_{im}\gamma_{jkln}^s=G^>_{jn}\gamma_{iklm}^s$, and similar for the second term. Therefore, various terms containing two-particle correlations can be rewritten in terms of fluctuations, i.e.,  $G^>_{im}\mathcal{G}_{jkln}$ or $G^>_{jn}\mathcal{G}_{iklm}$. This leads to four possibilities to approximate $\Gamma$, but only two of these will be considered here:
\begin{align}
    \Gamma_{ijklmn}^{(1)}&\coloneqq\pm G^>_{im}\gamma_{jkln}\pm G^<_{jl}\gamma_{ikmn},\label{eq:Gamma_P1} \\
    \Gamma_{ijklmn}^{(2)}&\coloneqq\pm G^>_{jn}\gamma_{iklm}\pm G^<_{km}\gamma_{ijln}. \label{eq:Gamma_P2}
\end{align}
Inserting equations~\eqref{eq:Gamma_P1} and~\eqref{eq:Gamma_P2} into equation~\eqref{eq:def-c3} leads to the following approximation for the three-particle term,
\begin{align}
    C^\mathrm{(\Gamma),PA}_{ijkl}\coloneqq&\pm\mathrm{i}\hbar\sum_{pqr}\left\{w_{ipqr}\Gamma^{(1)}_{rqjpkl}+w_{pjqr}\Gamma^{(2)}_{iqrkpl}\right\}  \nonumber\\
    &\mp\mathrm{i}\hbar\sum_{pqr}\left\{w_{pqkr}\Gamma^{(1)}_{irjpql}+w_{pqrl}\Gamma^{(2)}_{ijrkqp}\right\} \label{eq:C_polarization_approx} \\
    =&\pi^\mathrm{F}_{ijkl}+\left[U^{(2),\mathrm{F}},\gamma\right]_{ijkl}. \label{eq:C_P}
\end{align}
The commutator, $\left[U^{(2),\mathrm{F}},\gamma\right]$, contains the exchange contribution that is missing in the 2M approximation, equation~\eqref{eq:EOM_second_moment_approximation}. Together with the Hartree term in this equation, this yields the Hartree--Fock Hamiltonian,
\begin{equation}
\left[ h^{(2),\rm{H}},\gamma\right]_{ijkl}+\left[U^{(2),\mathrm{F}},\gamma\right]_{ijkl}=   \left[ h^{(2),\rm{HF}},\gamma\right]_{ijkl}. 
\end{equation}
Further, the term $\pi^\mathrm{F}$ in equation~\eqref{eq:C_P} is given by
\begin{equation}
    \pi^\mathrm{F}_{ijkl}\coloneqq\pi^\pm_{ijkl}-\pi_{ijkl},
\end{equation}
where $\pi^\pm$ is given by the replacement $w\rightarrow w^\pm$. This term thus denotes exchange contributions to the polarization term which is also present in the G1-G2 equations, cf. the term $\Pi^\pm$, equation~\eqref{eq:pi-def}. 

\subsubsection{Polarization and $GW$ approximations for $\gamma$} \label{sss:polarization_approximation_gamma}
Collecting all terms together, we can write down the EOM for $\gamma$:
\begin{equation}
    \mathrm{i}\hbar\partial_t\gamma^\mathrm{PA}_{ijkl}=\left[ h^{(2),\rm{HF}},\gamma^\mathrm{PA}\right]_{ijkl}+\pi^\pm_{ijkl}.
    \label{eq:gamma-pa}
\end{equation}
This is already a significant improvement compared to the 2M approximation,  equation~\eqref{eq:EOM_second_moment_approximation}, because we recovered the missing exchange corrections. 

Let us now investigate to what approximation in the G1-G2-scheme equation~\eqref{eq:gamma-pa} corresponds. Proceeding as in section~\ref{ss:approximation_of_second_moments}, the corresponding equation for $\mathcal{G}$ is found by adding the equations for $\gamma^{\rm PA}$ and $\gamma^{\rm PA,s}$, equation~\eqref{eq:EOM_gamma^s_P}, and introducing $\mathcal{G}^{\rm PA}\coloneqq \gamma^\mathrm{PA}\mp\gamma^{\mathrm{PA},s}$:
\begin{equation}
    \mathrm{i}\hbar\partial_t\mathcal{G}^\mathrm{PA}_{ijkl}=\left[h^{(2),\mathrm{HF}},\mathcal{G}^\mathrm{PA}\right]_{ijkl}+\Psi^\pm_{ijkl}+\Pi^\pm_{ijkl},
\end{equation}
where $\Psi^\pm$ and $\Pi^\pm$ are defined in equations~\eqref{eq:source_term_corr} and~\eqref{eq:pi-def}.
This equation can be considered the fluctuation equivalent of the $GW$ approximation with exchange corrections included, i.e., to $GW^\pm$. Thus, we conclude that the polarization approximation, $\gamma^{\rm PA}$, is equivalent to $GW^\pm$, as long as the weak coupling conditions i)--iii) are fulfilled. 

However, this is not a conserving approximation because complete (anti-)symmetrization requires to also include the particle-hole ladder (TPH) terms which leads to the replacement $\Pi^\pm \to P^\pm$, cf. equation~\eqref{eq:ppm-def}~\cite{joost_prb_22}. So far, the TPH term has not been identified in the fluctuation approach. Thus, the polarization approximation, equation~\eqref{eq:gamma-pa} is directly applicable to systems for which the exchange corrections vanish.

In order to derive the $GW$ approximation without exchange, we now look for modifications to equation~\eqref{eq:gamma-pa} which do not contain the exchange terms, i.e., $\Psi^\pm \to \Psi$ and $\Pi^\pm\to \Pi$. 
Such an equation for $\gamma$ can be indeed derived using condition iii), $|\mathcal{G}|\ll |\gamma^{(s)}|$, and additionally  neglecting the exchange contributions to the polarization term,
\begin{align}
    C^{(\Gamma)}_{ijkl}&\approx \left[U^{(2),\mathrm{F}},\gamma\right]_{ijkl},\\
    \pi^\mathrm{F}_{ijkl}&\rightarrow 0.
\end{align}
This leads to the following EOM for the two-particle fluctuation which we denote by $\gamma^{\mathrm{GW}}$:
\begin{equation}
    \mathrm{i}\hbar\partial_t \gamma^\mathrm{GW}_{ijkl}=\Big[ h^{(2),\HF},\gamma^\mathrm{GW}\Big]_{ijkl}+\pi_{ijkl},
    \label{eq:gamma-pa-nox}
\end{equation}
which, together with equation~\eqref{eq:EOM_gamma^s_P} for the source term, corresponds to the familiar $GW$-G2 equation
\begin{equation}
    \mathrm{i}\hbar\partial_t\mathcal{G}^\mathrm{GW}_{ijkl}=\Big[ h^{(2),\HF},\mathcal{G}^\mathrm{GW}\Big]_{ijkl}+\Psi_{ijkl}+\Pi_{ijkl}. \label{eq:G2_GW_approximation}
\end{equation}
The equivalence of the approximations of the fluctuation approach, equation~\eqref{eq:gamma-pa-nox} and the G1-G2 scheme on the $GW$ level holds for weak to moderate coupling, due to the assumptions i)--iii) and analogous approximations in the G1-G2 scheme (the neglect of TPP terms). To extend the fluctuation approach to strong coupling, one has to consider additional contributions from the residual term $R^{(\gamma)}$ [cf. equation~\eqref{eq:residual_term}] and from three-particle fluctuations, $\Gamma$. 

\subsubsection{Polarization and $GW$ approximations expressed in terms of the single-particle fluctuation $\delta \hat G$} \label{sss:polarization_approximation_1p}

Equivalently to approximating three-particle fluctuations, it is possible to derive the quantum polarization approximation solely on the basis of single-particle fluctuations. This can be done by again considering equation~\eqref{eq:EOM_deltaG}. In any approximation, it must be noted that the expectation value of single-particle fluctuations needs to be zero and thus also their time derivative, i.e., $\langle\partial_t\delta\hat{G}\rangle=\partial_t\langle\delta\hat{G}\rangle=0$. This means that any approximation of $[\delta\hat{U}^\mathrm{H},\delta\hat{G}]$ needs to also properly account for the collision integrals, $I^{(\gamma)}+I^{(\gamma)\dagger}$, in order to ensure this condition.

As it was seen in the derivation of the quantum polarization approximation, the missing contributions are given by the commutator of the effective two-particle Fock-potential with two-particle fluctuations and the exchange part of the polarization term. Similarly to the relation between correlations and fluctuations at the two-particle level [cf. equation~\eqref{eq:G2_fluc}], we now define an analogous relation for the fluctuation operator $\delta\hat{\gamma}$, where we take into account the linearity of equation~\eqref{eq:G2_fluc}
\begin{align}
    \delta\hat{\gamma}_{ijkl}&\equiv\delta\hat{G}_{ik}\delta\hat{G}_{jl}-\langle\delta\hat{G}_{ik}\delta\hat{G}_{jl}\rangle \nonumber\\
    &\equiv \pm \delta\hat{\gamma}^s_{ijkl}+\delta\hat{\mathcal{G}}_{ijkl}, \label{eq:fluc_G2_fluc}\\    
    \delta\hat{\gamma}^s_{ijkl} &\coloneqq 
    G^>_{il}\delta\hat{G}_{jk} + \delta\hat{G}_{il}G^<_{jk}.
\end{align}
Analogously to the derivation of the TDHF approximation, the quantum polarization approximation can be regarded as an approximation that neglects ``fluctuations'' of two-particle correlations, i.e., $\delta\hat{\mathcal{G}}_{ijkl}\equiv 0$, 
\begin{equation}
    \delta\hat{\gamma}^\mathrm{PA}_{ijkl}\coloneqq \pm \delta\hat{\gamma}^s_{ijkl} =\pm \left\{G^>_{il}\delta\hat{G}_{jk} + \delta\hat{G}_{il}G^<_{jk}\right\}.
\end{equation}
This leads to the following EOM for $\delta\hat{G}^\mathrm{PA}$,
\begin{equation}
 \mathrm{i}\hbar\partial_t\delta\hat{G}^\mathrm{PA}_{ij}=\left[h^\mathrm{HF},\delta\hat{G}^\mathrm{PA}\right]_{ij}+\left[\delta\hat{U}^\mathrm{HF},G^<\right]_{ij}.
 \label{eq:EOM-deltag}
\end{equation}
We now verify that this equation is equivalent to  EOM for $\gamma^\mathrm{PA}$, by differentiating the definition
\begin{equation}
    \mathrm{i}\hbar\partial_t\gamma^\mathrm{PA}_{ijkl}=\mathrm{i}\hbar\partial_t\left\langle \delta\hat{G}^\mathrm{PA}_{ik}\delta\hat{G}^\mathrm{PA}_{jl}\right\rangle,
\end{equation}
which contains the following terms,
\begin{align}
   &\underbrace{\bigg\langle\left[h^\mathrm{HF},\delta\hat{G}^\mathrm{PA}\right]_{ik}\delta\hat{G}^\mathrm{PA}_{jl}\bigg\rangle+\left\langle\delta\hat{G}^\mathrm{PA}_{ik}\left[h^\mathrm{HF},\delta\hat{G}^\mathrm{PA}\right]_{jl}\right\rangle}_{\left[ h^{(2),\rm{HF}},\gamma^\mathrm{PA}\right]_{ijkl}} \nonumber\\+& \underbrace{\left\langle\left[\delta\hat{U}^\mathrm{HF},G^<\right]_{ik}\delta\hat{G}^\mathrm{PA}_{jl}\right\rangle+\left\langle\delta\hat{G}^\mathrm{PA}_{ik}\left[\delta\hat{U}^\mathrm{HF},G^<\right]_{jl}\right\rangle}_{\pi_{ijkl}^\pm}.
\end{align}
We have thus proven that approximation \eqref{eq:EOM-deltag} is equivalent to the quantum polarization approximation for $\gamma^{\rm PA}$, equation~\eqref{eq:gamma-pa}. 
Recall that, to recover the polarization ($GW^\pm$) approximation  for $\mathcal{G}$, additionally, in the equation for $\gamma^s$ the $R$ term should be neglected, giving rise to $\gamma^{\rm PA, s}$, equation~\eqref{eq:EOM_gamma^s_P}.

In the same way, the standard $GW$ approximation can also be recovered. This is achieved by assuming $\delta \hat{\gamma}_{ijkl}\approx \pm G^<_{jk}\delta\hat{G}_{il}$,  leading to the EOM for $\fluc^\mathrm{GW}$,
\begin{equation}
    \mathrm{i}\hbar\partial_t \fluc^\mathrm{GW}_{ij}=\Big[ h^\HF,\fluc^\mathrm{GW}\Big]_{ij}+\Big[\delta\hat{U}^\hh,G^<\Big]_{ij}. \label{eq:EOM-deltag_GW}
\end{equation}
The two equations~\eqref{eq:EOM-deltag} and~\eqref{eq:EOM-deltag_GW} are remarkable results. They contain the complex physics corresponding to the nonequilibrium $GW$ approximation (with and without exchange, respectively) in an equation for a single-particle quantity $\delta \hat G$. This equation is closed and, obviously, much simpler than the corresponding pair of equations in the G1-G2-scheme or the corresponding KBE of two-particle NEGF theory.

The main problem, however, is that this EOM for the single-particle fluctuations is an operator equation, for which a direct solution is not possible. This problem can be solved approximately by using a semi-classical averaging approach. The result provides an approximation for $GW$ and $GW^\pm$, respectively, at the single-particle level, for weak to moderate coupling. This will be discussed more in detail in section~\ref{ss:stochastic_polarization_approximation}. However, it is not trivially possible to include contributions to compensate for the residual term [cf. equation~\eqref{eq:residual_term}] at the single-particle level. Thus, these approximations cannot be directly extended to the strong coupling regime in this framework at the moment.

\section{Stochastic mean-field theory} \label{s:SMF}
\subsection{General concept} \label{ss:SMF_general_concept}

The basic ideas of the stochastic mean-field theory (SMF)~\cite{lacroix_prb14} are the replacement of quantum mechanical operators by stochastic quantities (random realizations ``$\lambda$''), and of the quantum mechanical expectation value by an arithmetic mean,
\begin{align}
    \hat{A} &\longrightarrow A^\lambda,   \\
    \langle \hat{A}\rangle & \longrightarrow \overline{A^\lambda}.
    \label{eq:semiclassical-average}
\end{align}
In practice, random realizations ``$\lambda$'' of the initial state are generated, according to a statistical ensemble, and then propagated in time. This allows the operator equations to be solved approximately, and thus makes possible a solution of the many-body problem at the single-particle level, by sampling single-particle density matrices and propagating them using simple mean-field dynamics. Each realization of the initial state can now be propagated independently, thus reducing the solution of the many-body problem 
to a) the construction of the probability distribution describing the initial state, b) mean-field dynamics and c) semiclassical averaging~\cite{lacroix_prb14}. 

However, by replacing operators with random variables, important properties of operators are not properly accounted for, and the procedure requires special care. For example, the expectation value of products of two non-commuting operators, $\hat{A},\hat{B}$, should be symmetrized, i.e.,
\begin{equation}
    \hat{A}\hat{B}= \frac{1}{2}\left[\hat{A},\hat{B}\right]_++\frac{1}{2}\left[\hat{A},\hat{B}\right]_-.
\end{equation}
Now, the commutator can be  evaluated, and the symmetric expression (anti-commutator) can be replaced with the random ensemble, i.e., $[\hat{A},\hat{B}]_+/2\rightarrow A^\lambda B^\lambda$. This procedure is already known from other (semiclassical) approaches such as the symmetric representation of phase space operators~\cite{kubo64}. 

We now apply this procedure to our fluctuation approach and define symmetric two-particle fluctuations as
\begin{equation}
    \Tilde{\gamma}_{ijkl}\coloneqq \frac{\gamma_{ijkl}+\gamma_{jilk}}{2}= \frac{1}{2}\left\langle\left[\fluc_{ik},\fluc_{jl}\right]_+\right\rangle \label{eq:sym_gamma_def},
\end{equation}
and find, on the time-diagonal,
\begin{align}
    \gamma_{ijkl}&=\Tilde{\gamma}_{ijkl}+\frac{1}{2}\left\langle\left[\fluc_{ik},\fluc_{jl}\right]\right\rangle \nonumber \\
    &=\Tilde{\gamma}_{ijkl}\pm\frac{1}{2\mathrm{i}\hbar}\left\{\delta_{il}G^<_{jk}-\delta_{jk}G^<_{il}\right\}. \label{eq:gamma_sym-gamma_relation}
\end{align}
Similarly to fluctuation operators, semiclassical fluctuations are defined as deviations from the average,
\begin{equation}
    \Delta G^{\lambda}_{ij}\coloneqq G^{<,\lambda}_{ij}-\overline{G^{<,\lambda}_{ij}}\equiv G^{<,\lambda}_{ij}-\Tilde{G}^<_{ij}.
\end{equation}
We now apply this approach to equation~\eqref{eq:EOM_G_op} and solve an EOM for $G^{<,\lambda}$ or construct a system of differential equations for semiclassical fluctuations, $\Delta G^{<,\lambda}$, and $\Tilde{G}^<$. Analogously to the quantum fluctuation approach, one can now derive an EOM for $\overline{\Delta G^{\lambda}\Delta G^{\lambda}}$ that couples to semiclassical three-particle fluctuations, therefore leading to a hierarchy of coupled equations~\cite{Lacroix2016}. Thus, the solution of an EOM for $G^{<,\lambda}$ is equivalent to the solution of the entire hierarchy. While this can be considered as an advantage of the SMF theory, it is necessary to find a suitable probability distribution which correctly describes all statistical moments that are known from quantum fluctuations as well as their relation to correlations. This  will be discussed more in detail in section~\ref{ss:prob_dist}. Any probability distribution that fails to correctly reproduce all moments leads to the propagation of wrong contributions for higher order fluctuations. However, constructing the correct distribution can be shown to be impossible due to the differences of quantum mechanical and statistical expectation values~\cite{Lacroix2019}. 

\subsection{Observables}\label{ss:observables}
The SMF approach also allows one to calculate expectation values of observables. For an arbitrary one-body observable $A^{(1)}$, the expectation value can be computed as
\begin{equation}
    \big\langle \hat{A}^{(1)}\big\rangle=\pm\mathrm{i}\hbar\sum_{ij} A^{(1)}_{ij} G^{<}_{ji}. \label{eq:1p-observable}
\end{equation}
For a two-body observable $A^{(2)}$, we first have to consider the expectation value in the framework of $\gamma$,
\begin{align}
    \big\langle \hat{A}^{(2)}\big\rangle =-\hbar^2\sum_{ijkl}A^{(2)}_{ijkl}G^{(2),<}_{klij}=-\hbar^2\sum_{ijkl}A^{(2)}_{ijkl}\left(G^<_{ki}G^<_{lj}\mp\frac{1}{\mathrm{i}\hbar}\delta_{jk}G^<_{li}+\gamma_{klij}\right). \label{eq:2p-observable_gamma}
\end{align}
The expectation value of $A^{(2)}$ is then computed within the SMF approach by the symmetrized analog of equation~\eqref{eq:2p-observable_gamma}
\begin{align}
     \big\langle \hat{A}^{(2)}\big\rangle =-\hbar^2\sum_{ijkl}A^{(2)}_{ijkl}\left(G^<_{ki}G^{<}_{lj}\mp\frac{1}{2\mathrm{i}\hbar}(\delta_{jk}G^<_{li}+\delta_{li}G^<_{jk})+\overline{\Delta G^{\lambda}_{ki}\Delta G^{\lambda}_{lj}} \right). \label{eq:2p-observable}
\end{align} 
This can be generalized to an arbitrary $N$-particle observable.
\subsection{Probability distribution} \label{ss:prob_dist}
To construct the probability distribution describing the initial state at $t=t_0$, we consider an ideal (uncorrelated) state, i.e., $\mathcal{G}(t_0)=0$. We underline that this is not a restriction because, starting from an ideal initial state a correlated initial state, can be produced via the adiabatic switching protocol, e.g.,~\cite{schluenzen_jpcm_19}, which we also apply to the present fluctuation approach. This has the advantage that we only need to consider the properties of the initially ideal system and the corresponding probability distribution.

The expectation values of quantum fluctuation operators and their products are then given by 
\begin{align}
    \left\langle \fluc_{ij}(t_0)\right\rangle &=0,\label{eq:1st_moment}\\
    \left \langle \fluc_{ik}(t_0)\fluc_{jl}(t_0)\right\rangle &= -\frac{1}{\hbar^2}\delta_{il}\delta_{jk}n_i(1\pm n_j), \label{eq:2nd_moment}
\end{align}
with $n_i\coloneqq \pm\mathrm{i}\hbar G^<_{ij}(t_0)\delta_{ij}$. Applying the SMF approach to equations~\eqref{eq:1st_moment} and~\eqref{eq:2nd_moment}, we obtain
\begin{align}
    \overline{\Delta G^{\lambda}_{ij}(t_0)}&=0,\\
    \overline{\Delta G^{\lambda}_{ik}(t_0)\Delta  G^{\lambda}_{jl}(t_0)}&=-\frac{1}{2\hbar^2}\delta_{il}\delta_{jk}\{n_i(1\pm n_j)+n_j(1\pm n_i)\}.
\end{align}
Analogously, it is possible to derive expressions for higher moments of fluctuations, although this becomes increasingly more difficult~\cite{Lacroix2019}. The construction of a suitable probability distribution to best describe the initial state proves to be one of the main challenges of the SMF approach. The inability to construct a classical probability distribution can be seen when considering the second moment at zero temperature for fermions. Here, it follows 
\begin{equation}
   \overline{\Delta G^{\lambda}_{ik}(t_0)\Delta  G^{\lambda}_{jl}(t_0)}=-\frac{1}{2\hbar^2}\delta_{il}\delta_{jk}\overline{\delta}_{n_in_j},
\end{equation}
where we defined $\overline{\delta}_{ij}\coloneqq 1-\delta_{ij}$.
This means, in particular, that $\overline{|\Delta G^{\lambda}_{ii}|^2}=0$ and thus $\Delta G^{\lambda}_{ii}\equiv 0$, for all realizations.  However, the third and fourth moments are nonzero for expectation values that contain terms with matching indices, i.e., $\Delta G^{\lambda}_{ii}$. Further problems were highlighted in reference~\cite{Lacroix2019} when considering these higher moments. It was shown there that no standard probability distribution can simulate a quantum system exactly.

There exist a variety of approaches to approximately construct the initial state. Originally, standard probability distributions, e.g., a Gaussian distribution, are chosen so that the first moments are correctly reproduced without considering higher moments. This can be generalized by also considering quasiprobability distributions such as the Husimi distribution~\cite{Yilmaz2014}. We also explore Gaussian sampling, in section~\ref{ss:stochastic_sampling}. In addition, we introduce a new possibility to construct the initial state, which is not based on random sampling from a (quasi-)probability distribution but on a deterministic procedure. This is shown in section~\ref{ss:exact_sampling}.

\subsection{Dynamics} \label{ss:SMF_dynamics}
Since two-particle fluctuations appear only in the collision term, 
we now concentrate on the latter.
Inserting equation~\eqref{eq:gamma_sym-gamma_relation} into equation~\eqref{eq:collision_term_fluc} leads to [we drop the superscript $(\gamma)$]
\begin{equation}
    \Big[I+I^\dagger\Big]_{ij}=\Big[I^S+I^{S\dagger}\Big]_{ij}+\Big[S+S^\dagger\Big]_{ij}, \label{eq:sym_collision_term}
\end{equation}
where we used the definitions,
\begin{align}
    I^S_{ij}\coloneqq& \pm\mathrm{i}\hbar\sum_{klp}w_{iklp}\Tilde{\gamma}_{plkj}, \label{eq:collision_term_sym}\\
    S_{ij}\coloneqq& \frac{1}{2}\sum_{kl}w_{kljk}G^<_{il}. \label{eq:sym_contribution}
\end{align}
Therefore, using symmetric fluctuations, yields the following EOM for $G^<$:
\begin{align}
    \mathrm{i}\hbar\partial_tG^<_{ij}=&\Big[h^\mathrm{H},G^<\Big]_{ij}+\Big[S+S^\dagger\Big]_{ij}+\Big[I^S+I^{S\dagger}\Big]_{ij}. \label{eq:EOM_G1_sym}
\end{align}
Replacing $\Tilde{\gamma}$ in equation \eqref{eq:EOM_G1_sym} with $\overline{\Delta G^{\lambda}\Delta G^{\lambda}}$ results in the following equation for the  symmetrized  expectation value $\tilde G^<$:
\begin{align}
    \mathrm{i}\hbar\partial_t \Tilde{G}^{<}_{ij}=&\left[h^{\mathrm{H}},\Tilde{G}^{<}\right]_{ij}+\Big[S+S^\dagger\Big]_{ij}+\Big[I^\mathrm{SMF}+I^{\mathrm{SMF}\dagger} \Big]_{ij}, \label{eq:SMF-G<_fluc_sym}
\end{align}
with the SMF-collision term defined as 
\begin{equation}
    I^\mathrm{SMF}_{ij}\coloneqq\pm\mathrm{i}\hbar\sum_{klp}w_{iklp}\overline{\Delta G^{\lambda}_{pk}\Delta G^{\lambda}_{lj}}, \label{eq:collision_term_SMF}
\end{equation}
where the semi-classical averaging, equation~\eqref{eq:semiclassical-average}, is used.
Instead of using an additional equation of motion for the expectation value on the right and solving the entire fluctuation hierarchy, we can now proceed differently: we construct an EOM for $\Delta G^{\lambda}$ that is analogous to equation~\eqref{eq:EOM_deltaG} and solve it. The expectation value is computed at the final step.

To accomplish this task, we first construct an EOM for $G^{<,\lambda}$ which is itself symmetric:
\begin{equation}
    \mathrm{i}\hbar\partial_tG^{<,\lambda}_{ij}=\Big[h^{\mathrm{H},\lambda},G^{<,\lambda}\Big]_{ij}+\Big[S^\lambda+S^{\lambda\dagger}\Big]_{ij},
\end{equation}
where we introduced the definition
\begin{equation}
    S^\lambda_{ij}\coloneqq \frac{1}{2}\sum_{kl}w_{kljk}G^{<,\lambda}_{il}.
\end{equation} 
The EOM for $\Delta G^{<,\lambda}$ becomes,
\begin{align}
    \mathrm{i}\hbar\partial_t \Delta G^{\lambda}_{ij}=&\Big[h^\mathrm{H},\Delta G^{<,\lambda}\Big]_{ij}+\Big[\Delta U^\mathrm{H,\lambda},\Tilde{G}^<\Big]_{ij}+\Big[\Delta S^\lambda+\Delta S^{\lambda\dagger}\Big]_{ij}+\Big[\Delta U^\mathrm{H,\lambda},\Delta G^{<,\lambda}\Big]_{ij} \nonumber \\
    &-\Big[I^\mathrm{SMF}+I^{\mathrm{SMF}\dagger} \Big]_{ij},
\end{align}
where we used the definitions
\begin{align}
    \Delta S^\lambda_{ij}\coloneqq &\frac{1}{2}\sum_{kl}w_{kljk}\Delta G^{\lambda}_{il},\\
    \Delta U^\mathrm{H}_{ij}\coloneqq&\pm\mathrm{i}\hbar\sum_{kl}w_{ikjl}\Delta G^<_{lk}. \label{eq:fluc-hartree-potential_SMF}
\end{align}
When three-particle contributions are neglected, as in our case, $\Tilde{\gamma}$ obeys the same EOM as $\gamma$, due to the linearity of the equation. 

For completeness, we mention that, when symmetric three-particle contributions are taken into account, there arise differences in the equations of motion that are analogous to the case of the single-particle EOM. This leads to very complex equations, which are beyond the scope of the present work.

\subsection{Stochastic \textit{GW} approximation} \label{ss:stochastic_polarization_approximation}
The SMF approach can now be used to find new approximations for quantum fluctuations. Applying the SMF approach to the \textit{GW} approximation [cf. equation~\eqref{eq:EOM-deltag_GW}] allows for a treatment of this two-particle approximation at the single-particle level, therefore significantly reducing computational effort. 

Additionally, the neglect of any coupling to higher order fluctuations causes the neglect of contributions of higher moments, thus potentially minimizing the impact of choosing an approximate probability distribution. The stochastic \textit{GW} approximation (SGW) [stochastic version of equation~\eqref{eq:EOM-deltag_GW}] takes the following form:
\begin{align}
    \mathrm{i}\hbar\partial_t \Tilde{G}^{<,\mathrm{SGW}}_{ij}=&\Big[h^\mathrm{H},\Tilde{G}^{<,\mathrm{SGW}}\Big]_{ij}+\Big[S+S^\dagger\Big]_{ij}+\Big[I^\mathrm{SGW}+I^{\mathrm{SGW}\dagger}\Big]_{ij},\label{eq:EOM_SPA_G1} \\
    \mathrm{i}\hbar\partial_t \Delta G^{\mathrm{SGW},\lambda}_{ij}=&\Big[h^\mathrm{HF},\Delta G^{\mathrm{SGW},\lambda}\Big]_{ij}+\Big[\Delta U^{\mathrm{H,SGW},\lambda},G^{<,\mathrm{SGW}}\Big]_{ij}, \label{eq:EOM_SPA_deltaG}
\end{align}
where the following definitions were introduced,
\begin{align}
    I^\mathrm{SGW}_{ij}&\coloneqq\pm\mathrm{i}\hbar\sum_{klp}w_{iklp}\overline{\Delta G^{\mathrm{SGW},\lambda}_{pk}\Delta G^{\mathrm{SGW},\lambda}_{lj}}, \label{eq:collision_term_SPA}\\
    \Delta U^{\mathrm{H,SGW},\lambda}_{ij}&\coloneqq \pm\mathrm{i}\hbar\sum_{kl}w_{ikjl}\Delta G^{\mathrm{SGW},\lambda}_{lk}. \label{eq:hartree-fock-potential_SPA}
\end{align}
Instead of considering equation~\eqref{eq:EOM-deltag_GW}, we can also derive an analogous expression for equation~\eqref{eq:EOM-deltag} by replacing $\Delta U^{\mathrm{H,PA},\lambda}$ with $\Delta U^{\mathrm{HF,PA},\lambda}$. This gives rise to a set of equations that is equivalent to the quantum polarization approximation which will be called stochastic polarization approximation (SPA).  

The SGW fluctuation equations, \eqref{eq:EOM_SPA_G1} and \eqref{eq:EOM_SPA_deltaG}, as well as the SPA equations should be solved together with the EOM for the single-particle Green's function. The resulting system of equations has got the same complexity as standard TDHF equations where the numerical scaling (CPU time) of this approximation is mostly determined by the number of samples $K$. Therefore, the total numerical scaling of the SPA and SGW models is given by $\mathcal{O}(KN_b^4N_t)$, where $N_b$ is the number of basis states and $N_t$ is the number of time steps. It is interesting to compare this to the scaling of the $GW$ approximation within the G1-G2-scheme. Therein, the CPU-time scaling is ~\cite{joost_prb_20} of order $\mathcal{O}(N_b^6N_t)$. Thus, we find that the stochastic approximations are advantageous for (up to coefficients that may be different) $K < N_b^2$. The specific choice of sampling methods provides flexibility and allows for further optimization depending on the system.
\section{Sampling} \label{s:sampling}
Since it is shown to be impossible to construct a classical probability distribution in order to correctly describe the initial quantum state, the standard approaches to SMF only consider the known probability distributions such as a Gaussian distribution or a uniform distribution. In what follows, we discuss some of the possibilities to construct the initial state, though, we only consider a fermionic system at $T=0$ with a finite basis. 

\subsection{Stochastic sampling} \label{ss:stochastic_sampling}
The first approach we want to discuss is the standard approach of approximating the quantum state with a known probability distribution. For the construction of a suitable probability distribution, we define the variance of the distribution as
\begin{equation}
    \sigma_{ij}\coloneqq \frac{1}{2}\overline{\delta}_{n_i n_j}.
\end{equation}
In the original formulation of the SMF approach, a complex Gaussian distribution was assumed~\cite{Lacroix2019},
\begin{equation}
    P^N_{ij}(x,y) = \frac{1}{2\piup\sigma^2_{ij}}\exp\left(-\frac{x^2+y^2}{2\sigma^2_{ij}}\right),
\end{equation}
where $x,y$ denote the real and imaginary part and are independently sampled. This distribution correctly reproduces the first and second moments, but not the higher moments. An alternative is given by a complex generalization of the two-point distribution,
\begin{align}
    P^{2p}_{ij}(x,y)= &\frac{1}{4}\Big[ \delta(y)\{\delta(x+\sqrt{2}\sigma_{ij})+\delta(x-\sqrt{2}\sigma_{ij})\}+\delta(x)\{\delta(y+\sqrt{2}\sigma_{ij})+\delta(y-\sqrt{2}\sigma_{ij})\}\Big].
\end{align}
While this distribution also fails to correctly reproduce higher moments, it approximates them  better than the Gaussian distribution. In reference~\cite{Lacroix2019}, several distributions were compared for the exactly solvable Lipkin--Meshkov--Glick model, where it was shown that the two-point distribution had the best agreement with the exact solution. This further illustrates the importance of an appropriate distribution and motivates the construction of model distributions beyond the simple Gaussian and two-point assumptions. In reference~\cite{Yilmaz2014}, the possibility of a quasiprobability distribution was discussed and compared to previous approaches, where it was shown to further improve the accuracy of the SMF theory.

In practical application, instead of calculating the expectation value, one usually generates samples from the (quasi-)probability distribution and calculates the average over all $K$ samples, i.e.,
\begin{equation}
    \overline{A^\lambda} \approx \frac{1}{K}\sum_{\lambda =1}^{K} A^\lambda, \label{eq:expectation_approx}
\end{equation}
which becomes exact in the limit $K\rightarrow\infty$. Thus, to achieve the desired accuracy for the expectation value, the number $K$ should be chosen sufficiently large.

\subsection{Deterministic sampling} \label{ss:exact_sampling}
As mentioned in section~\ref{ss:stochastic_sampling}, in most practical applications, random realizations of the initial state are generated, and the expectation value of the ensemble is approximated by the average over all samples, cf. equation~\eqref{eq:semiclassical-average}. By contrast, the idea of the present sampling method is to construct realizations of the initial state so that the average is exactly equal to the expectation value. To this end, we  consider the following system of nonlinear equations
\begin{align}
    \sum_{\lambda =1}^{M} \cfluc_{ij} &= 0, \label{eq:exact_sampling_1st}\\
    \sum_{\lambda =1}^{M} \cfluc_{ik} \cfluc_{jl} &= \frac{M}{2}\delta_{il}\delta_{jk}\overline{\delta}_{n_in_j}, \label{eq:exact_sampling_2nd}
\end{align}
where we defined $\cfluc_{ij} \coloneqq -\mathrm{i}\hbar \Delta G^\lambda_{ij}(t_0)$. By finding a solution to these equations, we can construct an ensemble of the initial state that exactly fulfills the constraints given by the first two moments. Here, $M$ is a finite parameter that should be chosen such that the system of nonlinear equations has a solution and thus depends on the size of the basis. \\

In what follows, we consider a fermionic system with the spin configurations $\uparrow,\downarrow$, though this approach can be generalized to arbitrary spins. Additionally, we assume spin symmetry of the initial state, i.e., $n_i^\uparrow = n_i^\downarrow$. Let $N_b\in\mathbb{N}$ denote the size of the basis for one spin component and $N_p,N_h\in\mathbb{N}$ denote the number of occupied and unoccupied orbitals, respectively, for one spin component, with $N_p+N_h = N_b$. Without loss of generality, we can assume $n_i^\uparrow = n_i^\downarrow = 1$ for $i=1,\dots,N_p$ and $n_i^\uparrow=n_i^\downarrow = 0$ for $i=N_p+1,\dots,N_b$.

The goal of this approach is not to find all solutions of the system of equations given by \eqref{eq:exact_sampling_1st} and~\eqref{eq:exact_sampling_2nd}, for a given $M$, but rather to construct a single solution while minimizing the number of samples. The simplest way to fulfill equation~\eqref{eq:exact_sampling_1st} is to sample also $-\cfluc$, for every $\cfluc$, thus ensuring that the sum over all samples for every component is equal to zero. Additionally, equation~\eqref{eq:exact_sampling_2nd} can be partially fulfilled by setting $\cfluc_{ij}=0$ for $(i,j)\in \{1,\dots,N_p\}\times \{1,\dots,N_h\}\cup \{N_p+1,\dots,N_b\}\times \{N_h+1,\dots,N_b\}$. The matrix representing a single sample can then be written as
\begin{equation}
    \Delta n^{\lambda,\sigma} = \begin{pmatrix} 0 & \mathcal{A}^{\lambda,\sigma} \\ \mathcal{A}^{\lambda,\sigma\dagger} & 0\end{pmatrix}
\end{equation}
with $\mathcal{A}^{\lambda,\sigma} \in \CC^{N_h\times N_p}$. Here, it was used that $\cfluc$ is self-adjoint due to the properties of the single-particle density matrix. We can define a bijection\footnote{Any bijection $\varphi$ can be chosen here for the construction of the solution.} $\varphi:\,\{1,\dots,N_p\}\times \{1,\dots,N_h\}\rightarrow \{1,\dots,M\}$ with $M\coloneqq N_pN_h$. If we now set $K=8M$, we can find a solution to equations~\eqref{eq:exact_sampling_1st} and~\eqref{eq:exact_sampling_2nd} of the form 
\begin{align}
     \mathcal{A}_{ij}^{\alpha\beta,\uparrow} &= \begin{cases}\mathrm{i}^\beta M^{1/2},
     & \mbox{for } \varphi(i,j)=\alpha,\, \beta=1,\dots,4, \\ 0, & \mbox{else}, \end{cases}\\
        \mathcal{A}_{ij}^{\alpha\beta,\downarrow} &= \begin{cases}\mathrm{i}^\beta M^{1/2},& \mbox{for } \varphi(i,j)=\alpha,\, \beta=5,\dots,8, \\ 0, & \mbox{else}, \end{cases}
\end{align}
for $(\alpha,\beta)\in \{1,\dots, M\}\times\{1,\dots,8\}$. Using this approach for the construction of a solution, we require a total of $K=8N_pN_h$ samples, for each spin component, which means that this approach becomes numerically favorable for systems that are either mostly occupied or unoccupied. 

For example, for a two-state system with $N_p=N_h=1$, this leads to 16 samples of the form
\begin{align}
        \Delta n^{1,\uparrow}=\Delta n^{5,\downarrow}=-\Delta n^{2,\uparrow}=-\Delta n^{6,\downarrow} &= \begin{pmatrix} 0 & \mathrm{i} \\ -\mathrm{i} & 0\end{pmatrix},\\\Delta n^{3,\uparrow}=\Delta n^{7,\downarrow}=-\Delta n^{4,\uparrow}=-\Delta n^{8,\downarrow} &= \begin{pmatrix} 0 & 1 \\ 1 & 0\end{pmatrix}, \\
        \Delta n^{5,\uparrow}=\Delta n^{6,\uparrow}=\Delta n^{7,\uparrow}=\Delta n^{8,\uparrow}&= \begin{pmatrix} 0 & 0 \\ 0 & 0\end{pmatrix}, \\\Delta n^{1,\downarrow}=\Delta n^{2,\downarrow}=\Delta n^{3,\downarrow}=\Delta n^{4,\downarrow}&=\begin{pmatrix} 0 & 0 \\ 0 & 0\end{pmatrix}.
\end{align}
Although the first two moments are exactly reproduced, higher moments are not well approximated. In fact, the $n$-th moment can be proportional to $M^{({n-2})/{2}}$, for $n>2$, thus making this approach barely applicable to standard SMF theory. However, this approach turns out to be well suited for the SPA and SGW, because only the first two moments directly contribute here. Using this approach, we find that the total numerical scaling for SGW is given by $\mathcal{O}(N_pN_hN_b^4N_t)$. This sampling method leads to the same numerical scaling as for $GW$ if we consider a system with $N_p=N_h=N_b/2$. However, we find that for systems away from half-filling, this approach provides a significant improvement.

\section{Application to the Fermi--Hubbard model} \label{s:application_to_Hubbard}
In what follows, we consider the Hubbard model since it allows for simple numerical tests of our theoretical results. It is not only a key model of correlated electrons in condensed matter but also directly applicable to cold atoms in optical lattices. Furthermore, here extensive NEGF and G1-G2 results are available, e.g., references~\cite{schluenzen_prb16,schluenzen_prb17,schluenzen_prl_20} that can be used for comparison.

\subsection{Hubbard Hamiltonian} \label{ss:Hubbard_Hamiltonian}
For the Fermi--Hubbard model, the general pair-interaction of equation~\eqref{eq:generic_Hamiltonian} transforms into
\begin{equation}
    w_{ijkl}^{\alpha\beta\gamma\delta}=U\delta_{ij}\delta_{ik}\delta_{il}\delta_{\alpha\gamma}\delta_{\beta\delta}\overline{\delta}_{\alpha\beta},
\end{equation}
with the on-site interaction $U$ and the spin components denoted by greek indices. Additionally, the kinetic energy is replaced by a hopping Hamiltonian 
\begin{equation}
    h_{ij}=-\delta_{\langle i,j\rangle}J, \label{eq:hopping_hamiltonian}
\end{equation}
which describes nearest-neighbor hopping. The total Hamiltonian is then given by
\begin{equation}
    \hat{H}=-J\sum_{\langle i,j\rangle}\sum_{\sigma\in\{\uparrow,\downarrow\}}\hat{c}^\dagger_{i\sigma}\hat{c}_{j\sigma}+U\sum_i \hat{n}_i^\uparrow\hat{n}_i^\downarrow.
\end{equation}

\subsection{Implementation} \label{ss:implementation_SPA}
The EOMs for the single-particle Green's function in SGW, equations~\eqref{eq:EOM_SPA_G1} and~\eqref{eq:EOM_SPA_deltaG}, take the following form, where we drop all superscripts ``$\mathrm{SGW}$'' and ``$<$'', for simplicity, e.g., $G^{<,\mathrm{SGW}}_{ij}\equiv G_{ij}$, 
\begin{align}
    \mathrm{i}\hbar\partial_t G^{\sigma}_{ij}=&\Big[h^{\sigma},G^{\sigma}\Big]_{ij}+\Big[I+I^\dagger\Big]^\sigma_{ij},\label{eq:EOM_SPA_G1_Hubbard}\\
     \mathrm{i}\hbar\partial_t \Delta G^{\lambda,\sigma}_{ij}=&\Big[h^{\sigma},\Delta G^{\lambda,\sigma}\Big]_{ij}+\Big[\Delta U^{\lambda,\sigma},G^{\sigma}\Big]_{ij}, \label{eq:EOM_SPA_deltaG_Hubbard}
\end{align}
where the Hartree(--Fock) Hamiltonian and fluctuation Hartree-potential in equations~\eqref{eq:EOM_SPA_G1_Hubbard} and~\eqref{eq:EOM_SPA_deltaG_Hubbard} in the Hubbard basis become [cf. equations~\eqref{eq:hartree-fock-hamiltonian} and~\eqref{eq:hartree-fock-potential_SPA}]
\begin{align}
    h_{ij}^{\sigma}&\equiv h_{ij}^{\mathrm{HF},\sigma}\equiv  h_{ij}^{\mathrm{H},\sigma}=-\delta_{\langle i,j\rangle}J-\mathrm{i}\hbar\delta_{ij}U G^{{\overline{\sigma}}}_{ii}, \label{eq:hartree-fock-hamiltonian_Hubbard} \\
    \Delta U^{\lambda,\sigma}_{ij}&\equiv\Delta U_{ij}^{\hh,\lambda,\sigma}\equiv\Delta U_{ij}^{\mathrm{HF},\lambda,\sigma}=-\mathrm{i}\hbar\delta_{ij}U\Delta G^{\lambda,{\overline{\sigma}}}_{ii}. \label{eq:hartree-fock-potential_SPA_Hubbard}
\end{align}
Here, $\sigma=\uparrow(\downarrow)$ implies $\overline{\sigma}=\downarrow(\uparrow)$. 
This shows that all exchange contributions 
vanish, due to the specific choice of the pair-interaction so that the SGW is equivalent to SPA. 

The collision term in equation~\eqref{eq:EOM_SPA_G1_Hubbard} takes the form [cf. equation~\eqref{eq:collision_term_SPA}]
\begin{equation}
    I^{\sigma}_{ij}=-\mathrm{i}\hbar U \overline{\Delta G^{\lambda,{\overline{\sigma}}}_{ii}\Delta G^{\lambda,\sigma}_{ij}}, \label{eq:collision_term_SPA_Hubbard}
\end{equation}
and the contributions due to symmetrization take the form [cf. equation~\eqref{eq:sym_contribution}]
\begin{equation}
    S_{ij}^\sigma=\frac{1}{2}UG^{\sigma}_{ij}=- S_{ij}^{\dagger,\sigma}.
\end{equation}
Thus, symmetrization also does not lead to any additional contributions in the Hubbard basis. The initial state of the system in the natural orbital basis of $n(t_0)$ is chosen such that 
\begin{align}
    G_{ij}^\sigma(t_0) &= - \frac{1}{\mathrm{i}\hbar}\delta_{ij}n^\sigma_{i},\\
    \overline{\Delta G_{ij}^{\lambda,\sigma}(t_0)} &= 0,\\
    \overline{\Delta G^{\lambda,\sigma}_{ij}(t_0)\Delta  G^{\lambda,\sigma'}_{kl}(t_0)}&=-\frac{1}{2\hbar^2}\delta_{il}\delta_{jk}\delta_{\sigma\sigma'}\overline{\delta}_{n^\sigma_in_j^{\sigma'}}.\label{eq:initial_condition}
\end{align}
Depending on the configuration of the system, it becomes necessary to perform a transformation from the natural orbital basis to the Hubbard basis. This can be achieved by diagonalization of the Hamiltonian and the transformation of $G$ and $\Delta G^\lambda$ using its eigenvectors. 

In general, it is necessary to compute a nontrivial interacting ground state from which the externally driven dynamics start. Here, this is done using the so-called ``adiabatic switching method''~\cite{schluenzen_prb16} by replacing the on-site interaction $U$ with a time dependent interaction $U(t)$. Calculations start at $t_s$ with an uncorrelated ground state, with $U(t_s)=0$. The on-site interaction is increased monotonously and rather slowly such that at $t_0$ the system is in a fully correlated ground state with $U(t_0)=U$. 

The observables we consider are the densities on all sites, $n_i(t)$, and, in addition, the kinetic energy~($E_\mathrm{kin}$), Hartree--Fock ($E_\mathrm{HF}$) and correlation energies ($E_\mathrm{cor}$) which follow from equations~\eqref{eq:1p-observable} and~\eqref{eq:2p-observable},
\begin{align}
    E_\mathrm{kin}&=\mathrm{i}\hbar J\sum_{\langle i,j\rangle}\sum_\sigma G^\sigma_{ji},\qquad E_\mathrm{HF}= -\hbar^2U\sum_{i}G^\uparrow_{ii}G^\downarrow_{ii},\qquad
    E_\mathrm{cor}= -\hbar^2U\sum_{i}\overline{\Delta G^{\lambda,\uparrow}_{ii}\Delta G^{\lambda,\downarrow}_{ii}}.
\end{align}
In Appendix~\ref{app:energy-conservation} it is shown that the approximations presented in section~\ref{sss:polarization_approximation_gamma} conserve the total energy for the Fermi--Hubbard model. This directly implies that total energy conservation is also given for SGW/SPA due to the structure of the underlying set of equations.

\subsection{Numerical results}
\begin{figure}[p]
    \centering
    \includegraphics[width=0.7\textwidth]{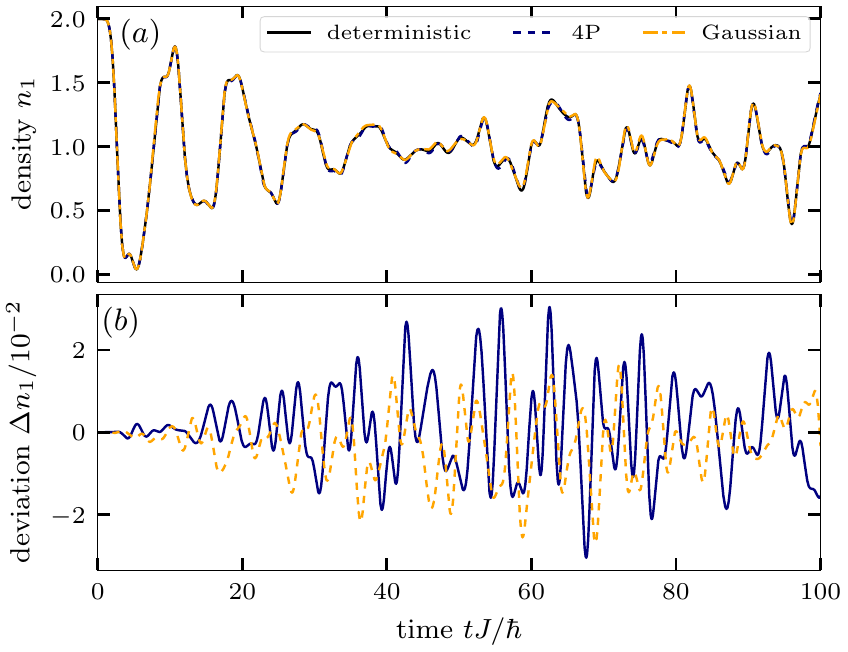}
    \caption{ (Colour online) Panel (a) corresponds to the density evolution of a half-filled $8$-site chain at $U=0.1J$ from SGW for initial states generated from deterministic sampling (section~\ref{ss:exact_sampling}) and random sampling (section~\ref{ss:stochastic_sampling}) using a complex four-point and Gaussian distribution with $10^4$ samples for each spin configuration. Starting point of the simulations was a noninteracting  initial state. The deviation displayed in (b) is computed as $\Delta n=n^\mathrm{det}-n^\mathrm{random}$.\label{fig:akbari_random_exact_0.1}}
    \vspace{5.5ex}
    \includegraphics[width=0.7\textwidth]{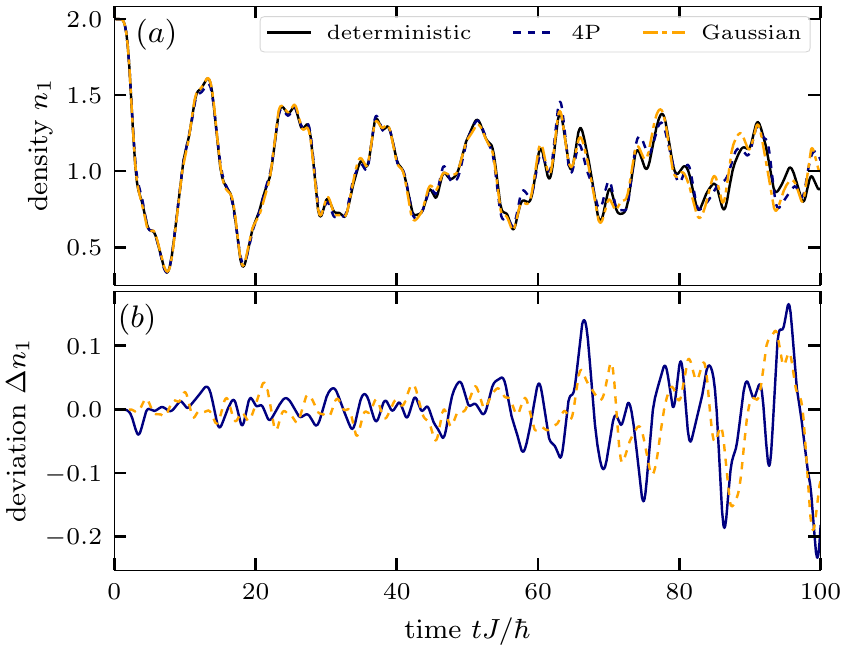}
    \caption{(Colour online) Same as figure~\ref{fig:akbari_random_exact_0.1}, but for $U/J=1.0$.\label{fig:akbari_random_exact_1.0}}
\end{figure}
\begin{figure}[t]
    \centering
    \includegraphics[width=0.85\textwidth]{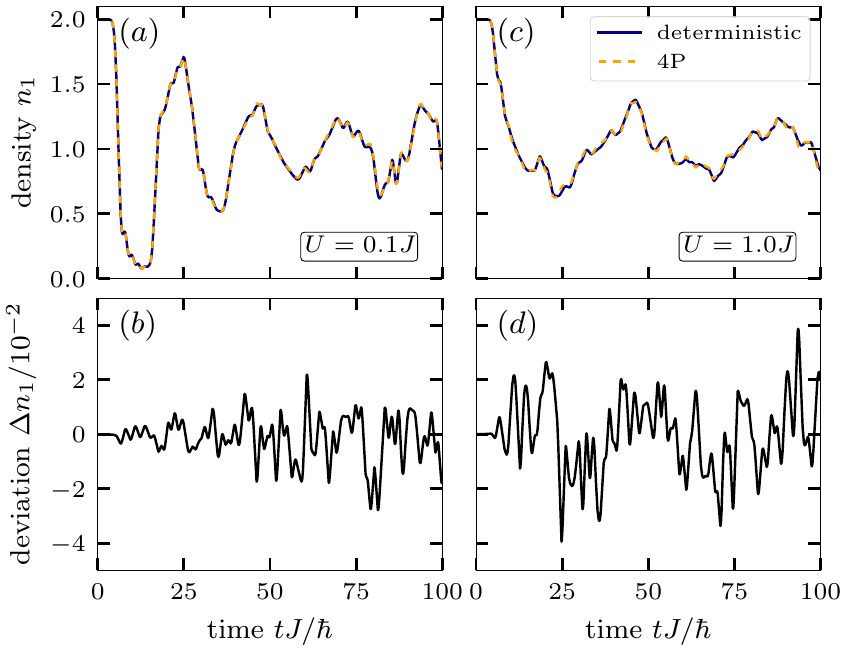}
    \caption{(Colour online) Density evolution of a half-filled $20$-site chain at $U=0.1J$ (a) and $U=1.0J$~(c) from SGW, for initial states generated from deterministic and random sampling using a complex four-point distribution with $10^4$ samples for each spin configuration. Starting point of the simulations was an uncorrelated initial state. The deviation for $U=0.1J$ (b) and $U=1.0J$ (d) is computed as $\Delta n =n^\mathrm{det}-n^\mathrm{random}$.}
    \label{fig:akbari_random_exact_ns20}
\end{figure}
\subsubsection{Comparison of sampling methods} \label{sss:comparison_sampling_methods}
First, we compare three different sampling methods. These are the complex four-point and Gaussian distributions, as well as the deterministic sampling method. In particular, we compare numerical results for 1D chains without periodic boundary conditions for different system sizes and interaction strengths. The test system is a chain with eight sites, of which the leftmost four sites are fully occupied and the right-hand half is empty, with a weak on-site interaction ($U=0.1J$). For the two stochastic sampling methods $10^4$ random realizations are used, for each spin component. The deterministic sampling method has $128$ samples for each spin component.

\begin{figure}[t]
    \centering
    \includegraphics[width=0.7\textwidth]{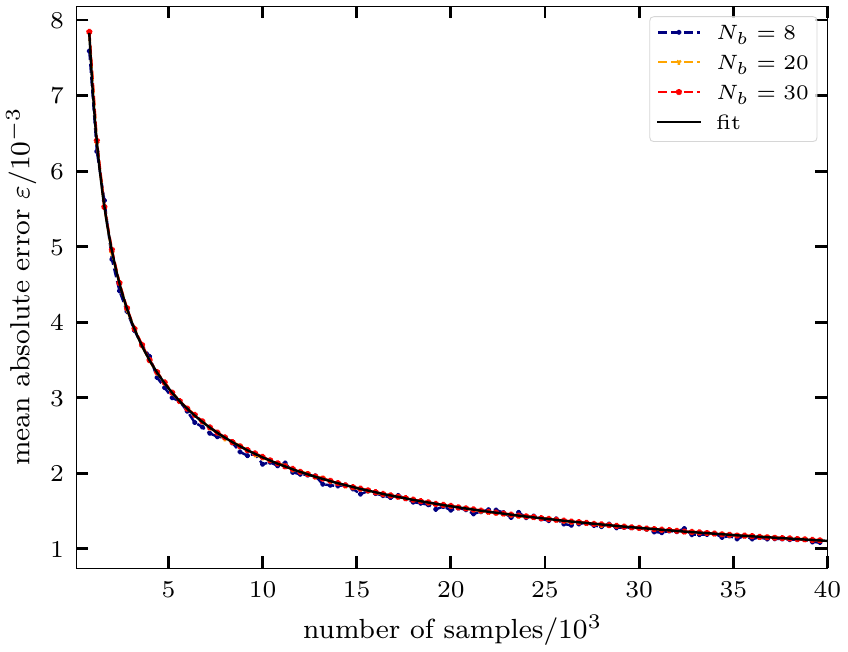}
    \caption{(Colour online) Mean absolute deviation $\varepsilon$ [cf. equation~\eqref{eq:mean_deviation}] for half-filled chains with $N_b=8,20,30$ using the complex four-point distribution and a fit, which is proportional to $1/\sqrt{K}$.}
    \label{fig:convergence}
\end{figure}

Figure~\ref{fig:akbari_random_exact_0.1} shows that all distributions display good agreement with each other. In particular, the distributions have absolute deviations of less than $0.02$ for all times, where the deviation is defined as the density difference between the deterministic sampling method and the  stochastic methods. Thus, the results of the deterministic sampling method can be interpreted as the limit of the dynamics of stochastic approaches and that all distributions with the same first two moments lead to the same dynamics. This is a significant difference compared to the standard SMF approach, since there the choice of the distribution for generating the initial state has a considerable influence on the dynamics of the system. 

Next, we extend the simulations for this system to an interaction strength of $U=1.0J$. Despite the same number of samples, we see in figure~\ref{fig:akbari_random_exact_1.0} significant deviations in the density dynamics for times $t\gtrsim 50\hbar/J$, showing that a stronger coupling leads to greater sensitivity with respect to deviations from the ideal initial state. Yet, for the cases considered, there does not appear to be any significant difference between the two stochastic sampling methods. 

Therefore, it is of interest to investigate this dependence on the interaction strength also for other systems. To this end, we consider a chain with twenty sites, where again the ten leftmost sites are fully occupied and the remaining ones are empty. Again, we consider the cases of $U=0.1J$ and $U=1.0J$ and use $10^4$ samples for each spin component for the stochastic sampling method. Here, we  only use the complex four-point distribution, since all considered probability distributions lead to equivalent results within this approximation. For the deterministic sampling method, we now use $800$ samples for each spin component. In figure~\ref{fig:akbari_random_exact_ns20} we see the behavior for the twenty-site chain similar to the behaviour of the eight-site chain for a weak on-site interaction: the absolute value of the deviation is smaller than $0.02$, for all times. Differences between the two system sizes are more apparent for an interaction strength of $U=1.0J$. Here, the density deviations are slightly larger than for $U=0.1J$, but they do not exceed 0.04 in magnitude. The dynamics of the two considered systems differ mainly in their speed. Deviations therefore seem to occur earlier the faster the dynamics proceed. This suggests that for a strong coupling, large systems are less sensitive to deviations from the exact initial fluctuations than smaller systems. Hence, the deterministic sampling method is advantageous over the stochastic one for small systems, especially for a stronger coupling. For the systems we considered, the required number of samples for a converged calculation was found to depend only weakly on the system size. This is further highlighted when considering the mean absolute error of the initial samples, which we define by 
\begin{equation}
    \varepsilon \coloneqq \frac{1}{4 N_b^4}\sum_{ijkl}\sum_{\sigma\sigma'} \left|\overline{\Delta G^{\lambda,\sigma}_{ik}(t_0)\Delta G^{\lambda,\sigma'}_{jl}(t_0)}-\Tilde{\gamma}^{\sigma\sigma'\sigma\sigma'}_{ijkl}(t_0)\right|. \label{eq:mean_deviation}
\end{equation} 
Using this expression, we can quantify the deviations from the ideal initial state given by equation~(\ref{eq:initial_condition}). Figure~\ref{fig:convergence} shows that the mean absolute error is nearly independent of the system size $N_b$. Additionally, our fit with a function proportional to $1/\sqrt{K}$ illustrates the expected trend due to the central limit theorem. Assuming $10^4$ samples for each spin component and half-filling so that $N_p=N_h=N_b/2$, stochastic sampling appears to be advantageous, especially for system sizes larger than $100$ sites. 

\subsubsection{Comparison of SGW to \emph{GW}-G1-G2 and exact results} \label{sss:comparison_approximations}
We now compare the results obtained from SGW with the exact solution using exact diagonalization and the $GW$ approximation of the G1-G2 scheme. All further calculations for SGW are done using the deterministic sampling method. We again consider 1D chains without periodic boundary conditions for different on-site interactions and systems sizes. 
\begin{figure}[p]
    \centering
   \includegraphics[width=0.7\columnwidth]{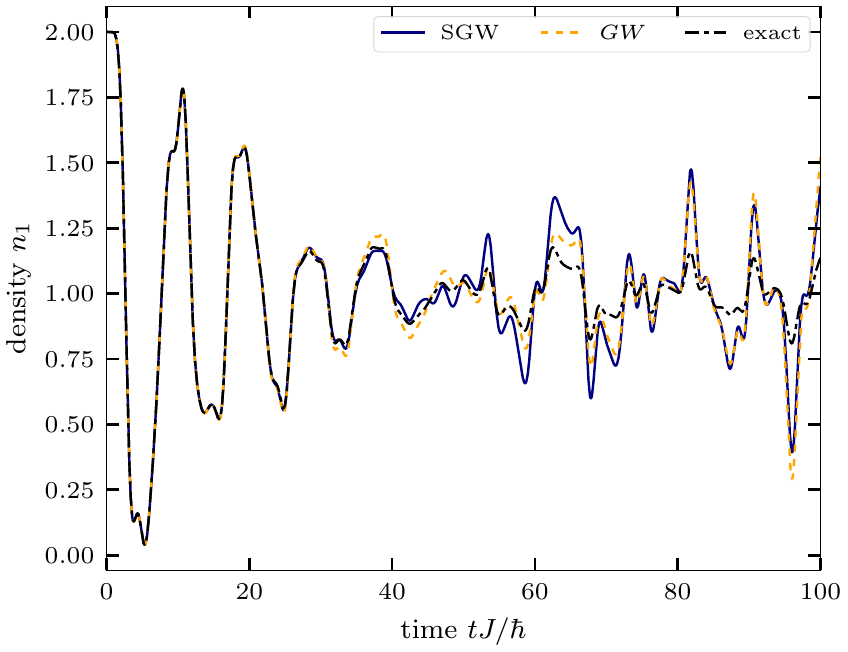}
    \caption{(Colour online) Evolution of the density from SGW, the $GW$-G1-G2 scheme and exact diagonalization for a half-filled $8$-site chain at $U=0.1J$. The initial state of the system was uncorrelated. \label{fig:akbari_Ns8_U0.1}}
    \vspace{5.5ex}
    \includegraphics[width=0.7\columnwidth]{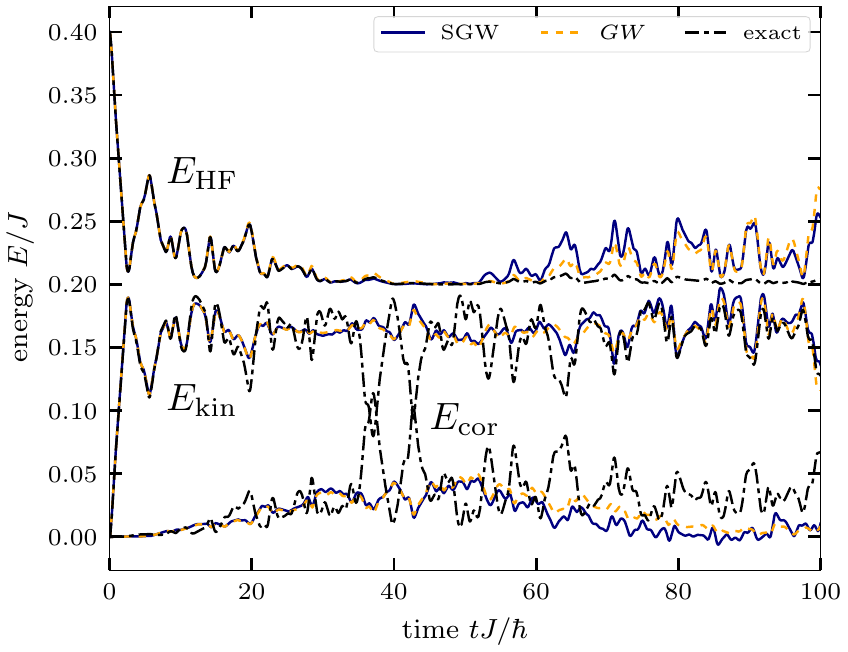}
    \caption{(Colour online) Evolution of the kinetic, HF and correlation energy from SGW, the $GW$-G1-G2 scheme and exact diagonalization for a half-filled $8$-site chain at $U=0.1J$. The initial state of the system was uncorrelated. \label{fig:akbari_Ns8_U0.1_energies}}
\end{figure}

\begin{figure}[p]
    \centering
    \includegraphics[width=0.7\columnwidth]{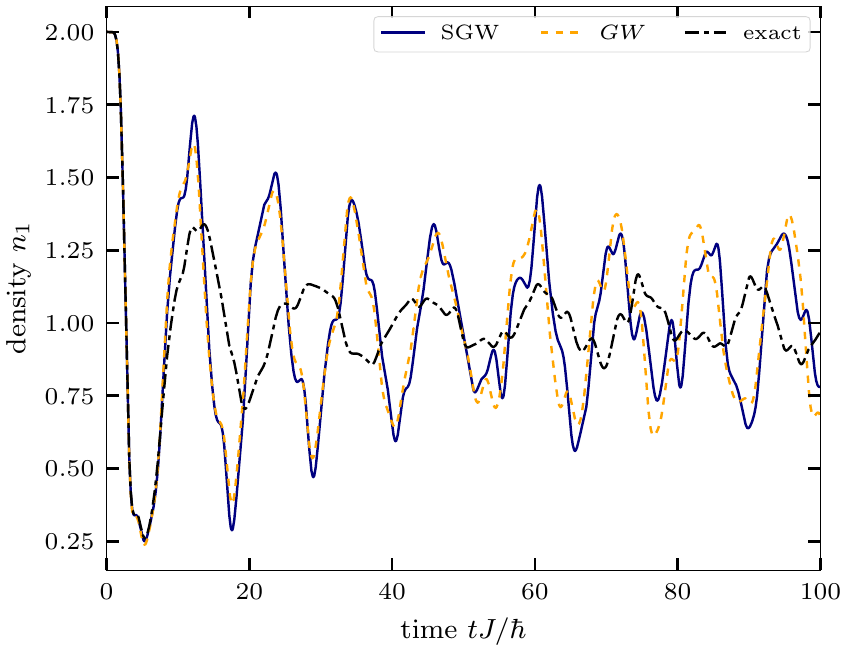}
    \caption{(Colour online) Same as figure~\ref{fig:akbari_Ns8_U0.1}, but for $U=0.5J$. \label{fig:akbari_Ns8_U0.5}}
 \vspace{5.5ex}
    \includegraphics[width=0.7\columnwidth]{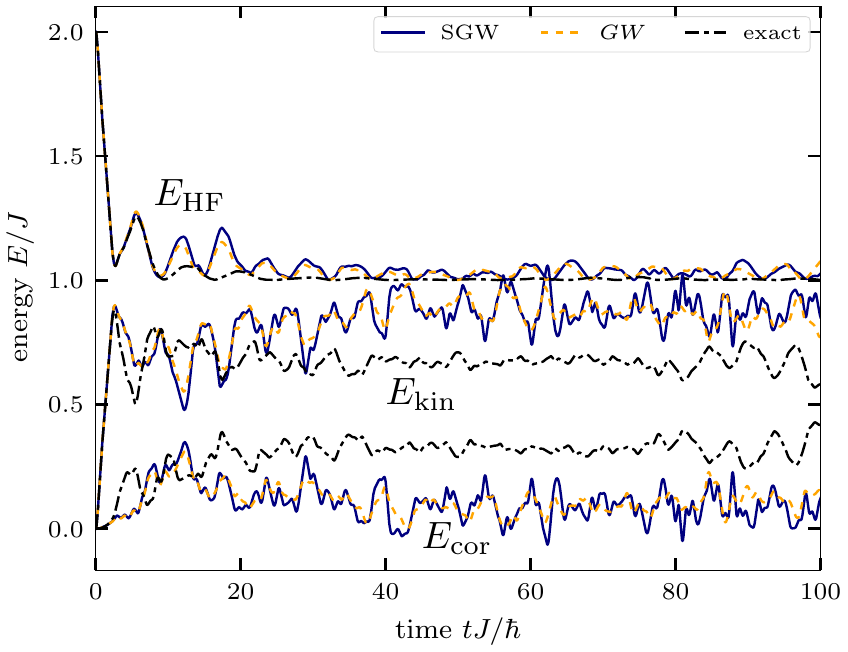}
    \caption{(Colour online) Evolution of the kinetic, HF and correlation energy from SGW, the $GW$-G1-G2 scheme and exact diagonalization for a half-filled $8$-site chain at $U=0.5J$. The initial state of the system was uncorrelated. \label{fig:akbari_Ns8_U0.5_energies}}
\end{figure}

\begin{figure}[t]
    \centering
    \includegraphics[width=0.7\columnwidth]{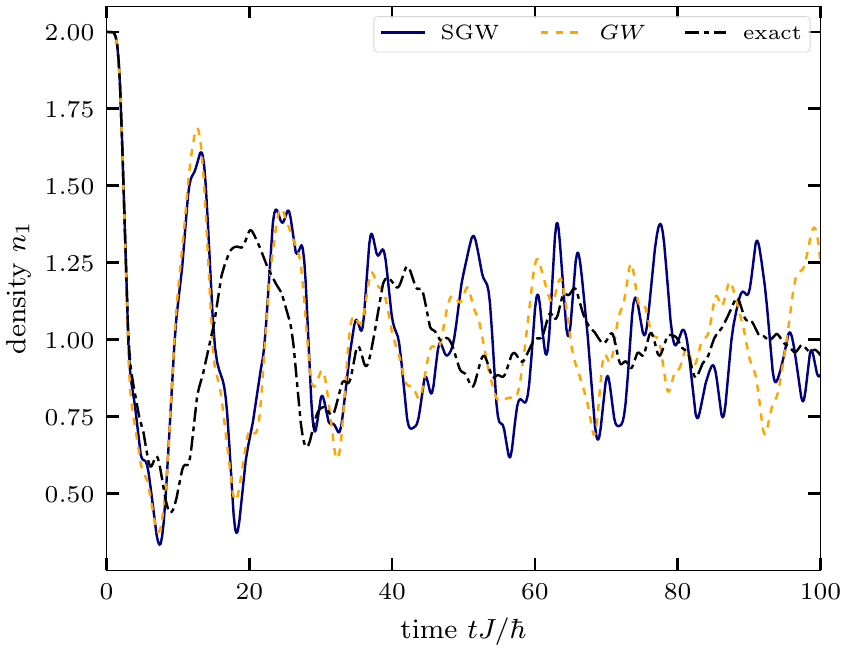}
    \caption{(Colour online) Same as figure~\ref{fig:akbari_Ns8_U0.1}, but for $U=1.0J$.}
    \label{fig:akbari_Ns8_U1.0/2.0}
\end{figure}

First, we consider a half-filled eight-site chain with weak on-site interaction ($U=0.1J$). The initial state is chosen so that the leftmost half of the sites is fully occupied. It can be seen in figure~\ref{fig:akbari_Ns8_U0.1} that the density dynamics  from SGW and the $GW$-G1-G2 scheme show good agreement with each other. Both reproduce the exact result reasonably well for times $t\lesssim 40\hbar/J$, after which SGW qualitatively displays a better agreement for times $t\lesssim 50\hbar/J$. After this point, deviations between the two approximations and the exact result start to increase, with the $GW$ approximation showing slightly better density dynamics than SGW with respect to the exact solution. However, this trend is only observable at the beginning and, already for times $t\lesssim 70\hbar/J$, there are only minimal deviations between the two approximations, although both fail to correctly reproduce the exact dynamics of the system. While the exact dynamics displays some sort of damping behavior with increasing time, we observe a revival of the dynamics for both approximations which is due to the missing damping in the HF propagators of the GKBA approach.

Next, consider the energy dynamics of this system. As seen in figure~\ref{fig:akbari_Ns8_U0.1_energies} the same overall trend as for the density dynamics can be observed. However, the different energies display a different behavior, i.e., the HF energy shows good agreement for all calculations for times $t\lesssim 50\hbar/J$, after which a revival of the dynamics can be observed for SGW and $GW$. Both approximations show a qualitative good agreement and only small deviations arise. Similar results can be seen when considering the kinetic energy. However, already for times $t\gtrsim 20\hbar/J$, there are significant differences between the two approximations and the exact result. However, SGW and $GW$ show a very good agreement between each other, for the entire time propagation. For times $t\gtrsim70\hbar/J$, we can see a good agreement for all three calculations. Finally, consider the correlation energy. Here, we observe the same overall behavior as for the kinetic energy, although SGW and $GW$ show deviations for later times that compensate the difference in HF energy. In conclusion, the energy dynamics shows an analogous behavior to the density dynamics. Both SGW and $GW$ fail to correctly reproduce the kinetic and correlation energy by significantly underestimating correlations. Additionally, we see that SGW conserves the total energy within the accuracy of the numerical calculation, thus confirming the results presented in Appendix~\ref{app:energy-conservation}.

\begin{figure}[p]
    \centering
   \includegraphics[width=0.7\columnwidth]{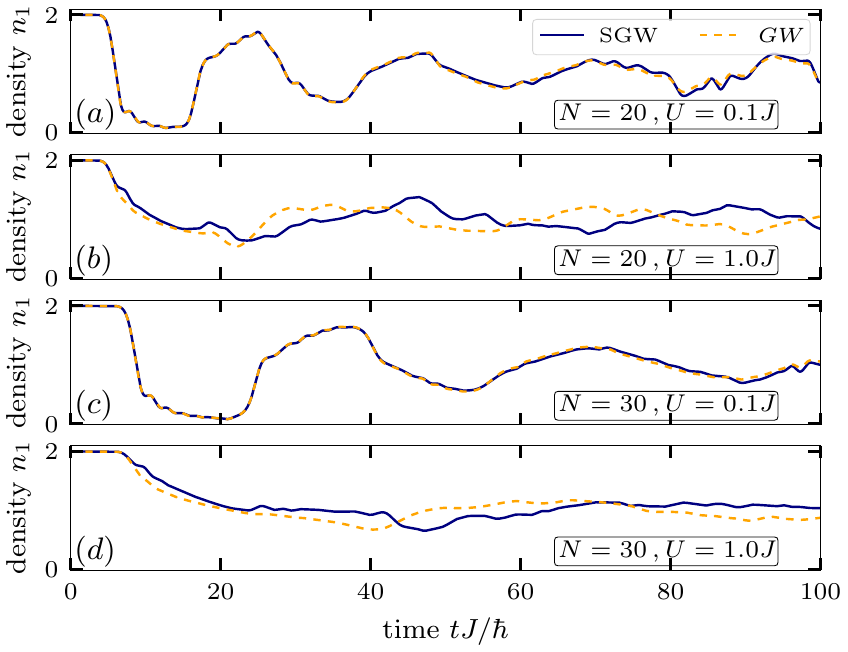}
    \caption{(Colour online) Density evolution of a half-filled 20-site chain at $U=0.1J$ (a) and $U=1.0J$~(b) from SGW and the $GW$-G1-G2 scheme. (c) and (d) show the density evolution for a 30-site chain for $U=0.1J$ and $U=1.0J$, respectively. Starting point of the simulations was a uncorrelated initial state. \label{fig:akbari_Ns20/30_U0.1/1.0}}
     \vspace{5.5ex}
   \includegraphics[width=0.75\columnwidth]{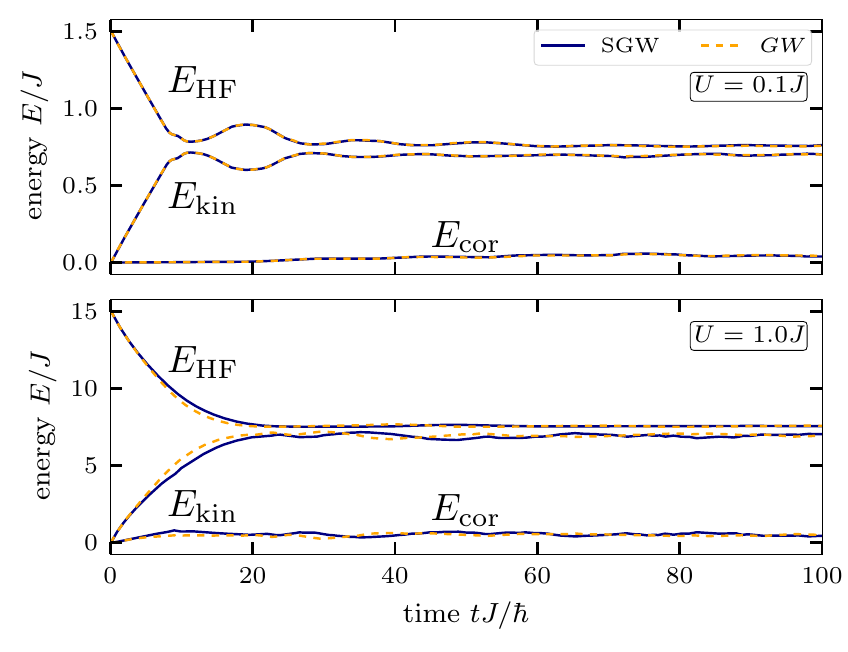}
    \caption{(Colour online) Evolution of the kinetic, HF and correlation energy from SGW and the $GW$-G1-G2 scheme for a half-filled 30-site chain at $U=0.1J$ (a) and $U=1.0J$ (b). The initial state of the system was uncorrelated. \label{fig:akbari_Ns30_energies}}
\end{figure}

\begin{figure}[p]
    \centering
    \includegraphics[width=0.7\columnwidth]{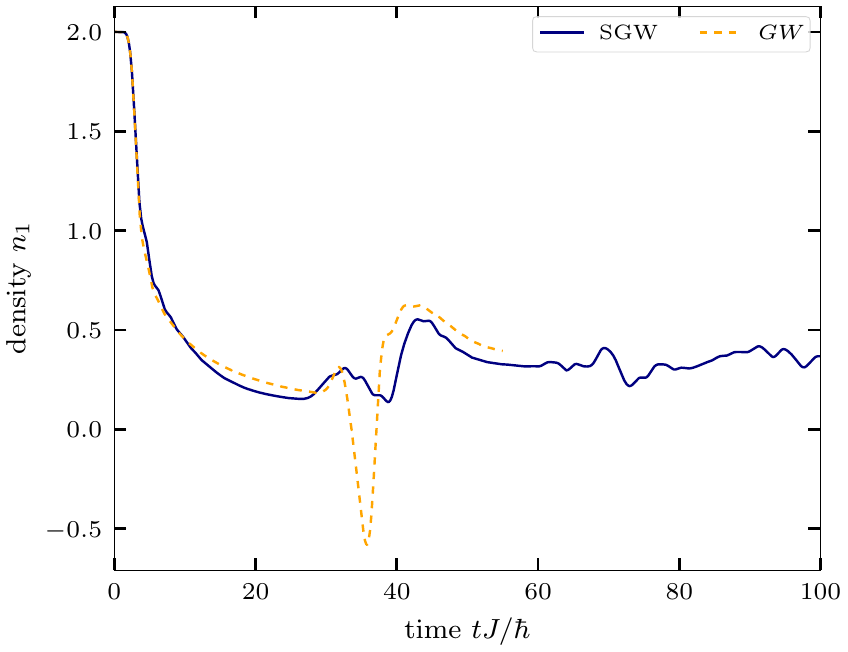}
    \caption{(Colour online) Density dynamics on the first site of a 30-site chain with the leftmost five sites fully occupied for $U=1.0J$ from SGW and the $GW$-G1-G2 scheme. Starting point of the calculations was an uncorrelated state. The $GW$ results diverge for $t\geqslant 55 \hbar/J$.\label{fig:ns30-u1}}
     \vspace{5.5ex}
    \includegraphics[width=0.7\columnwidth]{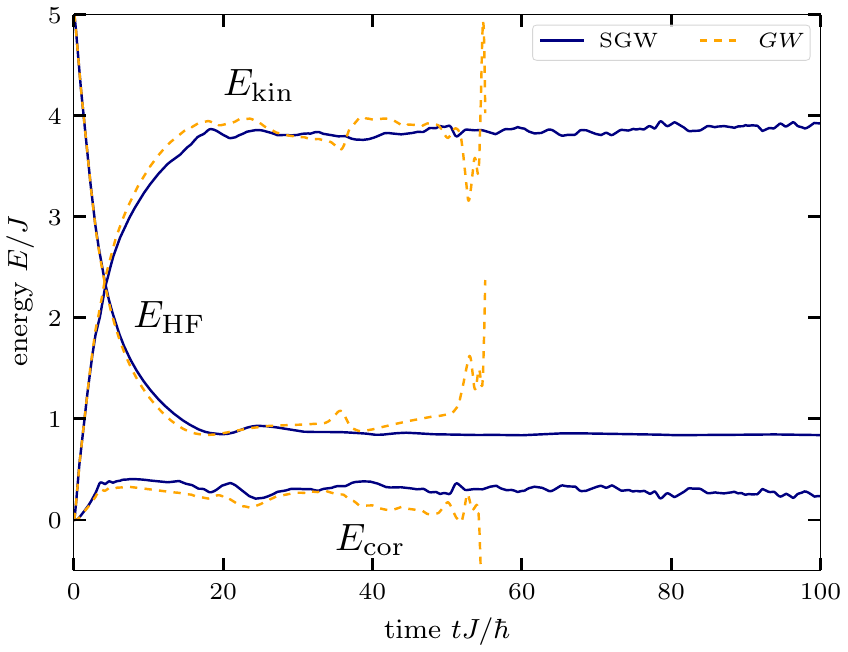}
    \caption{(Colour online) Dynamics of the kinetic, HF and correlation energies of the setup described in figure~\ref{fig:ns30-u1}. \label{fig:ns30-u1-energy}}
    
\end{figure}
\begin{figure}[t]
    \centering
    \includegraphics[width=0.6\columnwidth]{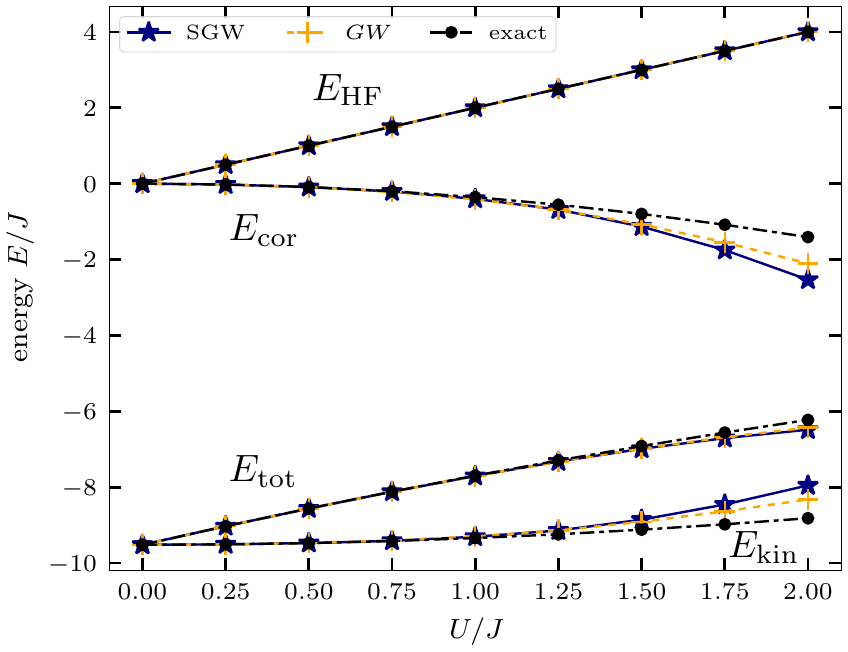}
    \caption{(Colour online) Ground state calculations of the kinetic, HF, correlation and total energies for $U=0.0J,\dots,2.0J$ for a half-filled $8$-site chain. SGW and $GW$ energies are calculated using the adiabatic switching method and the exact energies by direct calculation of the correlated ground state using exact diagonalization.}
    \label{fig:ground_state_energies}
\end{figure}

Next, we consider the same system but with an increased  on-site interaction of $U=0.5J$. In figure~\ref{fig:akbari_Ns8_U0.5} we see that the SGW and $GW$ show a very good agreement between each other for times $t\lesssim 40 \hbar/J$, whereas after this time small differences show up. For times $t\gtrsim 60\hbar/J$, the oscillations of the density for both approximations have a similar amplitude. At the same time, both approximations reproduce the exact dynamics only for $t \lesssim 10 \hbar$. For larger times, the amplitude of the oscillations has the correct order of magnitude, but the dominant frequencies are not captured. Again, the deviations of SGW and $GW$ from the exact benchmark are comparable.

Figure~\ref{fig:akbari_Ns8_U0.5_energies} shows that the Hartree--Fock energy of the SGW and $GW$ agree well with each other. For the kinetic and correlation energies, we observe deviations similar to the density dynamics. Again, both approximations significantly underestimate the correlations and thus overestimate the kinetic energy of the system.

One of the main assumptions in the derivation of SGW (as well as the $GW$ approximation) was that correlations are small. It is therefore interesting to consider the density dynamics of a system for stronger coupling, i.e., $U\geqslant 1.0J$, in order to estimate the range of validity. In figure~\ref{fig:akbari_Ns8_U1.0/2.0} we again consider a half-filled eight-site chain where the leftmost sites are fully occupied. For $U=1.0J$, we observe significant deviations of both approximations compared to the exact results, and the deviations start earlier when the coupling is increased, in agreement with earlier stochastic mean field calculations~\cite{lacroix_prb14}. Interestingly, the overall trend of the density --- relaxation towards $1.0$ at long times --- is correctly reproduced by SGW and $GW$.
Additionally, we observe increasing differences between SGW and $GW$ with increasing coupling. For $U=1.0J$, both approximations show a good agreement for times $t\lesssim 40\hbar/J$.

Finally, let us compare SGW to $GW$ for larger systems with $20$ and $30$ sites, even though we do not have exact benchmarks available. We again consider an initial state where the leftmost half of the sites is completely filled and the remaining sites are empty. In figure~\ref{fig:akbari_Ns20/30_U0.1/1.0}, we find that for $U=0.1J$, both approximations show a good agreement between each other for $N=20$ as well as $N=30$, showing that for a slower dynamics in larger systems and weak coupling the equivalence of the SGW and $GW$ holds for longer times. The agreement of the density dynamics between both approximations, for $U=1.0J$, now extends to longer times, i.e., $t\sim 40\hbar/J$ ($t\sim 20\hbar/J$), for $30$ ($20$) sites. At the same time, figure~\ref{fig:akbari_Ns30_energies} shows that the energy dynamics are in very good agreement, between both approximations, not only for $U=0.1J$ but even for $U=1.0J$. Likewise, in figures~\ref{fig:ns30-u1} and~\ref{fig:ns30-u1-energy} the density and energy dynamics of SGW and $GW$ are compared for a thirty-site chain with $U=1.0J$ where in the initial state only the first five sites are fully occupied. Up to a time of $t\sim 50\hbar/J$, both methods are in good agreement except for a short period around $t\sim 35\hbar/J$ where $GW$ shows an unphysical negative density. Further, at $t\sim 55\hbar/J$, the $GW$ dynamics become unstable which is not reproduced by SGW. Instability is a well known issue of the time-linear approach. In order to fix it, in general one has to resort to the numerically expensive procedure of purification~\cite{joost_prb_22}. Therefore, the increased numerical stability of SGW compared to $GW$ is a huge advantage of the stoachastic approach.

After comparing SGW and $GW$ for systems far from equilibrium, we
now inquire how well ground state properties agree with each other. To this end, we again consider a half-filled eight-site chain. We consider half filling, i.e., each site is singly occupied. Using the adiabatic switching method,  we calculate the ground state energies for varying on-site interaction strength for SGW and $GW$. As can be seen in figure~\ref{fig:ground_state_energies}, the HF energies for SGW, $GW$ agree with exact diagonalization results over the entire $U$ range. Slightly worse is the agreement of the total energies. On the other hand, kinetic and correlation energies are found to be in good agreement with the benchmark up to $U/J \sim 1.25$. 
As observed before, cf.~figure~\ref{fig:akbari_Ns8_U0.1_energies}, $GW$ displays slightly smaller deviations from the exact result than SGW.

\section{Discussion and outlook}\label{s:discussion}
In this paper we have developed a fluctuation approach to quantum many-particle systems in and out of equilibrium. Our approach was motivated by the classical fluctuation approach of Klimontovich~\cite{klimontovich_1982}. Indeed, our theory contains Klimontovich's results as a limiting case. In particular, the classical limit of the two-particle fluctuations, cf. equation~\eqref{eq:G2_fluc}, is directly transformed into the correlation function of the fluctuations of the microscopic phase space density $N(x) $\cite{klimontovich_1982},
\begin{align}
    \gamma_{ijkl} & \equiv \mathcal{G}_{ijkl} \pm G^>_{il}G^<_{jk} \nonumber\\
    &\to \mathcal{G}^{\rm cl}_{ijkl} + \delta_{il}\,n_{jk} \nonumber\\
    & \to n^2 g_2(x,x') + \delta(x-x') n f(x) \equiv \langle \delta N(x)\delta N(x')\rangle,
\nonumber
\end{align}
where, in the second line we used the (anti-)commutation rules of the field operators,  $G^>_{ij}=\frac{1}{\mathrm{i}\hbar}(\delta_{ij}\pm n_{ij})\approx \frac{1}{\mathrm{i}\hbar}\delta_{ij}$, and, in the third line, transformed from a general basis to the Wigner representation with $x=\textbf{r},\textbf{p}$. The second terms in the expression for the fluctuations are the respective source terms, $\gamma^s$, and their classical limit. A similar correspondence holds between the different approximations. In particular, our result for $\gamma^{\rm GW}$, which is equivalent to the $GW$ approximation in the classical limit, goes over to the non-Markovian Balescu--Lenard equation. While in the classical case this equation has never been solved numerically, in this paper we included extensive numerical results for the more general quantum case.

On the other hand, our work has been motivated by the stochastic mean field concept of Ayik and Lacroix. Our theory was formulated at the most fundamental level --- fluctuations of the single-particle Green's functions operators $\delta \hat G$. This allowed us to establish the connections between fluctuations and correlations and to derive the first equations of the fluctuations hierarchy. The equations of this hierarchy are formally simpler than the BBGKY hierarchy, containing fewer terms. However, a number of physical effects, such as exchange diagrams and strong coupling contributions, become entangled with other terms in a non-trivial manner. The reason is that the two-particle fluctuations, $\gamma$, contain contributions from both two-particle correlations, $\mathcal{G}$, and particle-hole fluctuations, $\gamma^s$, cf. equation~\eqref{eq:G2_fluc}.

In order to decouple the fluctuation hierarchy, we considered various approximations, in particular those that were introduced before, for classical systems, such as the approximations of first and second moments~\cite{klimontovich_1982}. While the approximation of first moments --- corresponding to the time-dependent Hartree (quantum Vlasov) approximation --- has a clear range of applicability, the approximation of second moments does not. We, therefore, investigated additional approximations and worked out which of them correspond to the known many-body approximations of the BBGKY or NEGF formalisms.
Of particular interest was to identify the $GW$ approximation with and without exchange corrections because it is a cornerstone for many-body theory for weakly and moderately coupled systems, both in the ground state and out of equilibrium. Here, we were able to identify the polarization approximation, $\gamma \to \gamma^{\rm PA}$, and the $GW$ approximation, $\gamma \to \gamma^{\rm GW}$, as the proper equivalent to the $GW$-G1-G2 equations with and without exchange, respectively. Both of them are valid for weak to moderate coupling such that the residual term $R^{(\mathcal{G})}$ could be neglected.
%

While the effort to solve the fluctuation equations (propagate $G_1$ and $\gamma$) is comparable to that of the G1-G2-scheme, the fluctuation approach has a large potential advantage: the computationally costly  propagation of two-particle quantity $\gamma$ can be avoided entirely if the corresponding equation of motion for~$\delta \hat G$ is solved instead.
%
Here, we were able to derive the equation of motion for $\delta \hat G$ that is equivalent [again, up to the term $R^{(\mathcal{G})}$] to the polarization approximation and, thus, to the $GW$ approximation with and without exchange corrections for weak and moderate coupling. Applying the concept of the semiclassical SMF approach, we mapped this problem on the stochastic PA (SPA) and stochastic $GW$ (SGW) models that can be solved by averaging over a large number $K$ of trajectories that start from different initial states. The numerical scaling of SGW vs $GW$-G1-G2 was found to be $\mathcal{O}(K N_b^4N_t)$ vs 
$\mathcal{O}(N_b^6N_t)$ which will be advantageous if the number of samples that is required is below $N_b^2$, i.e., for large systems. This is exactly the case where the G1-G2-scheme becomes computationally challenging, so the fluctuation approach gives access to situations that are presently out of reach with other methods.

We tested the SGW model for small Hubbard clusters (here SPA and SGW coincide) in the ground state and far from equilibrium. Moreover, we explored a variety of stochastic sampling methods including Gaussian sampling, the four point approach and a deterministic method. Our simulations confirm that
  all sampling methods that reproduce the first two moments correctly are equivalent, in the framework of the SGW approximation. Therefore, there is a particularly large freedom to minimize the numerical effort by an appropriate choice of the sampling method. The deterministic sampling method provides advantages especially for systems which are away from half filling. On the other hand, stochastic sampling is favorable for large systems (100 sites or larger), due to the weak dependence of the number of samples~$K$ on the system size $N_b$. In that case, the advantage of the stochastic approach, that scales in the Hubbard case as $\mathcal{O}(K N_b^2N_t)$, over the $GW$-G1-G2 scheme [$\mathcal{O}(N_b^4N_t)$] is maximal.

Our numerical SGW results showed excellent agreement with $GW$-G1-G2 results, and both, overall, agree well with the exact results, where available, within the applicability range of the approximation, i.e., for moderate coupling, $U/J \lesssim 1$. This confirms that the fluctuation approach constitutes an interesting and efficient alternative to NEGF or density operator methods, provided the relevant physical approximation has been identified. To extend the applicability range of the fluctuation approach in the future, it would be particularly interesting to  identify the strong coupling ($T$-matrix) approximation, e.g.,~\cite{schluenzen_cpp16}, as well as the dynamically screened ladder approximation~\cite{joost_prb_22}.

Finally, three fundamental issues concerning the limitations of our fluctuation approach should be addressed. The first issue is the validity range of replacing the quantum-mechanical average by a semiclassical mean, cf. equation~\eqref{eq:semiclassical-average}. This procedure is only exact if there are no quantum coherence effects between different samples. This is the case if the individual samples are selected for an ideal system, as was done in our simulations, or in a basis of natural orbitals. In that case, the density matrix is diagonal and no quantum coherence effects occur.
The second issue is the use of moments of the probability distribution that correspond to a non-interacting state. However, this is not an approximation if correlations are subsequently selfconsistently created via the adiabatic switching procedure, as in our approach. 
The third issue is the use of an ensemble of fluctuations of the initial state alone without any update of the ensemble at later times. Even if the density matrix is diagonal in the initial state, non-diagonal elements (quantum coherences) may develop in the course of the dynamics.
This is an issue in the standard stochastic mean field approach, e.g.,~\cite{lacroix_prb14} where an ensemble of realizations of the  single-particle Green's function operator, $\hat G_1$, is propagated and added semiclassically. However, our SGW approach is different: instead of many realizations of $G_1$ we propagate its expectation value quantum-mechanically. The main difference to the G1-G2 scheme lies  in the treatment of the collision term $I$. While in the G1-G2 scheme, $I$ is computed from a trace over the quantum two-particle correlation function $\mathcal{G}$, cf. equation~\eqref{eq:collision_term_corr}, in our fluctuations approach we compute the trace over $\overline{\Delta G^{\mathrm{SGW},\lambda}_{pk}\Delta G^{\mathrm{SGW},\lambda}_{lj}}$, cf. equation~\eqref{eq:collision_term_SPA}. This latter expression, in general, does not capture all quantum coherence effects that build up during the evolution. However, the computation of the trace gives rise to dephasing effects that tend to reduce the differences to the exact result. 

Fortunately, we are able to perform the ultimate test of the relevance of these three issues. In fact, our extensive numerical results that compare SGW and $GW$-G1-G2 simulations are an independent strong confirmation that no systematic errors are introduced by our stochastic approach for the systems and parameters under consideration. This gives strong confirmation for the power of the present fluctuation approach which should be very valuable, in particular for studying the dynamics of large quantum systems.

\section*{Acknowledgements}
We acknowledge helpful comments from Stefan Kehrein and Niclas Schlünzen. This work is supported by the Deutsche Forschungsgemeinschaft via project BO1366/16.

\appendix

\section{Three-particle correlations and fluctuations} \label{app:3p_corr_fluc_relation_derivation}
The relation between three-particle correlations, $\mathcal{G}^{(3)}$ and fluctuations, $\Gamma$, can be found in similar fashion as for two-particle correlations and fluctuations in equation~\eqref{eq:G2_fluc}. First, the full three-particle Green's function is  expressed in terms of fluctuations (all quantities depend on a single time),
\begin{align}
    G^{(3),<}_{ijklmn}=&\pm\frac{1}{(\mathrm{i}\hbar)^3}\langle\hat{c}^\dagger_l\hat{c}^\dagger_m\hat{c}^\dagger_n\hat{c}_k\hat{c}_j\hat{c}_i\rangle=\pm\frac{1}{(\mathrm{i}\hbar)^3}\left\{\pm\langle\hat{c}^\dagger_l\hat{c}^\dagger_m\hat{c}^\dagger_n\hat{c}_i\hat{c}_j\hat{c}_k\rangle\right\} \nonumber \\
    =&\pm\frac{1}{(\mathrm{i}\hbar)^3}\left\{\pm\langle\hat{c}^\dagger_l\hat{c}^\dagger_m(\mp\delta_{in}\pm\hat{c}_i\hat{c}^\dagger_n)\hat{c}_j\hat{c}_k\rangle\right\} \nonumber \\
    =&\pm\frac{1}{(\mathrm{i}\hbar)^3}\Big\{-\delta_{in}\langle\hat{c}^\dagger_l(\mp\delta_{jm}\pm\hat{c}_j\hat{c}^\dagger_m)\hat{c}_k\rangle+\langle\hat{c}^\dagger_l(\mp\delta_{im}\pm\hat{c}_i\hat{c}^\dagger_m)(\mp\delta_{jn}\pm\hat{c}_j\hat{c}^\dagger_n)\hat{c}_k\rangle\Big\} \nonumber \\
    =&\pm\frac{1}{(\mathrm{i}\hbar)^2}\delta_{jm}\delta_{in}G^<_{kl}-\frac{1}{\mathrm{i}\hbar}\delta_{in}(G^<_{jl}G^<_{km}+\gamma_{jklm}) \nonumber+\frac{1}{(\mathrm{i}\hbar)^2}\delta_{im}\delta_{jn}G^<_{kl}\mp\frac{1}{\mathrm{i}\hbar}\delta_{im}(G^<_{jl}G^<_{kn}+\gamma_{jkln})\nonumber\\ &\mp\frac{1}{\mathrm{i}\hbar}\delta_{jn}(G^<_{il}G^<_{km}+\gamma_{iklm})+G^<_{il}G^<_{jm}G^<_{kn}+G^<_{il}\gamma_{jkmn}+G^<_{jm}\gamma_{ikln}+G^<_{kn}\gamma_{ijlm}+\Gamma_{ijklmn} \nonumber \\
    =&\pm G^>_{in}G^>_{jm}G^<_{kl}\mp G^>_{in}G^<_{kl}G^<_{jm}\mp G^>_{jm}G^<_{in}G^<_{kl}\nonumber\\
    &\pm G^<_{in}G^<_{kl}G^<_{jm}-G^>_{in}G^<_{jl}G^<_{km}+G^<_{in}G^<_{jl}G^<_{km}\nonumber\\
    &+G^>_{jn}G^>_{im}G^<_{kl}-G^>_{jn}G^<_{im}G^<_{kl}-G^>_{im}G^<_{jn}G^<_{kl}\nonumber\\
    &+G^<_{kl}G^<_{im}G^<_{jn}\mp G^>_{im}G^<_{jl}G^<_{kn} \pm G^<_{im}G^<_{jl}G^<_{kn} \nonumber\\
    &\mp G^>_{jn}G^<_{il}G^<_{km} \pm G^<_{jn}G^<_{il}G^<_{km} +G^<_{il}G^<_{jm}G^<_{kn}\nonumber\\
    &-G^>_{in}\gamma_{jklm}+G^<_{in}\gamma_{jklm}\mp G^>_{im}\gamma_{jkln}\nonumber\\
    &\pm G^<_{im}\gamma_{jkln}\mp G^>_{jn}\gamma_{iklm}\pm G^<_{jn}\gamma_{iklm} \nonumber\\
    & + G^<_{im}\gamma_{jkln}+ G^<_{jm}\gamma_{ikln}+G^<_{kn}\gamma_{ijlm} \nonumber\\
    &+\Gamma_{ijklmn}. \label{eq:fullG3_gamma}
\end{align}
On the other hand, the three-particle Green's function is expressed via the cluster expansion by
\begin{align}
    G^{(3),<}_{ijklmn}=&G^<_{il}G^<_{jm}G^<_{kn}+G^<_{im}G^<_{jn}G^<_{kl}+G^<_{in}G^<_{jl}G^<_{km}\nonumber\\
    &\pm G^<_{im}G^<_{jl}G^<_{kn}\pm G^<_{in}G^<_{jm}G^<_{kl}\pm G^<_{il}G^<_{jn}G^<_{km}\nonumber \\
    &+G^<_{il}\mathcal{G}_{jkmn}\pm G^<_{im}\mathcal{G}_{jkln}\pm G^<_{in}\mathcal{G}_{jkml}\nonumber\\
    &+G^<_{jm}\mathcal{G}_{ikln}\pm G^<_{jl}\mathcal{G}_{ikmn}\pm G^<_{jn}\mathcal{G}_{iklm}\nonumber\\
    &+G^<_{kn}\mathcal{G}_{ijlm}\pm G^<_{kl}\mathcal{G}_{ijnm}\pm G^<_{km}\mathcal{G}_{ijln}\nonumber\\
    &+\mathcal{G}^{(3)}_{ijklmn}. \label{eq:G3_cluster}
\end{align}
Comparing equations~\eqref{eq:fullG3_gamma} and~\eqref{eq:G3_cluster}, we identify the general connection between the correlated three-particle Green's function and the three-particle and two-particle fluctuations:
\begin{align}
    \mathcal{G}^{(3)}_{ijklmn} = &\Gamma_{ijklmn}-G^>_{im}G^>_{jn}G^<_{kl}-G^>_{in}G^<_{jl}G^<_{km}\nonumber\\ \nonumber
    &\mp G^<_{jl}\mathcal{G}_{ikmn} \mp G^<_{km}\mathcal{G}_{ijln}- G^<_{kl}\mathcal{G}_{ijmn}\\ 
    &\mp G^>_{im}\mathcal{G}_{jkln}\mp G^>_{jn}\mathcal{G}_{iklm}-G^>_{in}\mathcal{G}_{jklm}  
    \\
    =&\Gamma_{ijklmn}-G^>_{in}G^<_{km}G^<_{jl}-G^>_{jn}G^>_{im}G^<_{kl} \nonumber \\
    &\mp G^>_{in}G^<_{jm}G^<_{kl}\mp G^>_{in}G^>_{jm}G^<_{kl} \nonumber\\
    &\pm G^<_{jl}\gamma_{ikmn} \pm G^<_{km}\gamma_{ijln}+ G^<_{kl}\gamma_{ijmn}\nonumber\\ 
    &\pm G^>_{im}\gamma_{jkln}\pm G^>_{jn}\gamma_{iklm}+G^>_{in}\gamma_{jklm}.
\end{align}

\section{Total energy conservation of the SGW approximation} \label{app:energy-conservation}
The total energy of a system is given by the sum of the single-particle and two-particle contributions [cf. equations~\eqref{eq:1p-observable} and~\eqref{eq:2p-observable_gamma}]. We start with a representation in terms of two-particle correlations, rather than fluctuations, so that it follows 
\begin{align}
    E_\mathrm{tot}=\pm\mathrm{i}\hbar\sum_{ij}h_{ij}G^<_{ji}-\frac{\hbar^2}{2}\sum_{ijkl}w_{ijkl}\left( G^{\mathrm{HF},(2),<}_{klij}+\mathcal{G}_{klij}\right), \label{eq:total_energy}
\end{align}
where $G^{\mathrm{HF},(2),<}$ denotes the Hartree--Fock contribution to the two-particle Green's function [cf. equation~\eqref{eq:G2_cluster}]. Differentiating equation~\eqref{eq:total_energy} with respect to time leads to
\begin{align}
    \frac{\mathrm{d}}{\mathrm{d}t}E_\mathrm{tot}=&\pm\sum_{ij}h_{ij}\left(\left[h^\mathrm{HF},G^<\right]_{ji}+\left[I^{(\mathcal{G})}+I^{(\mathcal{G})\dagger}\right]_{ji}\right)+\mathrm{i}\hbar\sum_{ijkl}w^\pm_{ijkl}G^<_{ki}\left(\left[h^\mathrm{HF},G^<\right]_{lj}+\left[I^{(\mathcal{G})}+I^{(\mathcal{G})\dagger}\right]_{lj}\right) \nonumber\\
    &-\frac{\hbar^2}{2}\sum_{ijkl}w_{ijkl}\frac{\mathrm{d}}{\mathrm{d}t}\mathcal{G}_{klij},
\end{align}
where we used that $w_{ijkl}G^{\mathrm{HF},(2),<}_{klij}=w^\pm_{ijkl}G^<_{ki}G^<_{lj}$ and that, using the exchange symmetries of the interaction tensor, allows us to rewrite the following derivative as 
\begin{equation}
    \sum_{ijkl}w^\pm_{ijkl}\frac{\mathrm{d}}{\mathrm{d}t}\left(G^<_{ki}G^<_{lj}\right)=2\sum_{ijkl}w^\pm_{ijkl}G^<_{ki}\frac{\mathrm{d}}{\mathrm{d}t}G^<_{lj}.
\end{equation}
Obviously, total energy conservation depends on the applied approximation of two-particle correlations (or fluctuations) and on the underlying system. For the \textit{GW} approximation, as given in section~\ref{sss:polarization_approximation_gamma} [cf. equation~\eqref{eq:G2_GW_approximation}, with included residual term $R$, cf. equation~\eqref{eq:residual_term}] and for the case of the Fermi--Hubbard system it follows, due to spin symmetry,
\begin{align}
    \frac{\mathrm{d}}{\mathrm{d}t}E_\mathrm{tot}=&\,2J\sum_{\langle i,j\rangle}\left(\left[h^\uparrow,G^{<,\uparrow}\right]_{ji}+\left[I^{(\mathcal{G})}+I^{(\mathcal{G})\dagger}\right]^\uparrow_{ji}\right)+2\mathrm{i}\hbar U\sum_{i}G^{<,\uparrow}_{ii}\left(\left[h^\downarrow,G^{<,\downarrow}\right]_{ii}+\left[I^{(\mathcal{G})}+I^{(\mathcal{G})\dagger}\right]^{\downarrow}_{ii}\right) \nonumber\\
    &-\mathrm{i}\hbar U\sum_{i}\left(\left[h^{(2),\uparrow\downarrow},\mathcal{G}^{\uparrow\downarrow\uparrow\downarrow}\right]_{iiii}+\Psi^{\uparrow\downarrow\uparrow\downarrow}_{iiii}+\Pi^{\uparrow\downarrow\uparrow\downarrow}_{iiii}-R^{(\mathcal{G}),\uparrow\downarrow\uparrow\downarrow}_{iii}\right).
\end{align}
In reference~\cite{joost_prb_22} it was shown that the \textit{GW}-G1-G2 approximation is energy conserving so that 
\begin{align}
    \frac{\mathrm{d}}{\mathrm{d}t}E_\mathrm{tot}=\mathrm{i}\hbar U\sum_iR^{(\mathcal{G}),\uparrow\downarrow\uparrow\downarrow}_{iiii}.
\end{align}
By inserting the definition of the residual term [cf. equation~\eqref{eq:residual_term}] and using that $G^{\lessgtr,\uparrow\downarrow}\equiv 0$, we find 
\begin{equation}
    R^{(\mathcal{G}),\uparrow\downarrow\uparrow\downarrow}_{iiii}=G^{>,\uparrow\downarrow}_{ii}\left[I^{(\mathcal{G})}+I^{(\mathcal{G})\dagger}\right]^{\downarrow\uparrow}_{ii}+G^{<,\downarrow\uparrow}_{ii}\left[I^{(\mathcal{G})}+I^{(\mathcal{G})\dagger}\right]^{\uparrow\downarrow}_{ii}\equiv 0.
\end{equation}
Therefore, $\frac{\mathrm{d}}{\mathrm{d}t}E_\mathrm{tot}\equiv 0$, so the total energy is conserved for the SGW approximation.


\newpage

\ukrainianpart

\title{Квантово-флуктуаційний підхід до нерівноважного \emph{GW} наближення}

\author {Е. Шрьоедтер, Я. -Ф. Йоост, М. Боніц}
\address{Інститут теоретичної фізики та астрофізики, Кільський університет ім. Крістіана Альбрехта, D-24098 Кіль, Німеччина}

\makeukrtitle

\begin{abstract}
	\tolerance=3000%
	
	Квантову динаміку ферміонних та бозонних багаточастинкових систем під дією зовнішнього збудження можна успішно вивчати за допомогою методів нерівноважних функцій Ґріна (НФҐ) або приведеної матриці густини.
	Апроксимації вводяться шляхом правильного вибору багаточастинкової власної енергії або розчепленням ланцюжка рівнянь ББҐКІ. Ці наближення ґрунтуються на методі діаграм Фейнмана або на кластерних розвиненнях в одночастинкових та кореляційних операторах.	Ми застосовуємо інший підхід, де замість рівнянь руху для багаточастинкових НФҐ (або операторів густини) аналізуються рівняння для кореляційних функцій флуктуацій.
	Ми отримуємо перші два рівняння альтернативної ієрархії флуктуацій та обговорюємо можливі наближення для розчеплення цих рівнянь. Зокрема, отримано поляризаційне наближення (ПН), яке є еквівалентним до нерівноважного \emph{GW} наближення з обмінними ефектами теорії НФҐ у межах узагальненого замикання Каданова-Бейма для випадку слабкого зв'язку. Основна перевага підходу квантових флуктуацій полягає в тому, що стандартне усереднення за ансамблем можна замінити напівкласичним середнім за різними початковими реалізаціями, як це було показано раніше Лакруа та співавторами [див. D. Lacroix \textit{et al.}, Phys. Rev. B, 2014, \textbf{90}, 125112]. Ми виконуємо стохастичне \emph{GW} наближення (SGW) а також наближення стохастичної поляризації (НСП), які в границі слабкого зв'язку, як показано в цій статті, є еквівалентними до \emph{GW} апроксимації як з врахуванням обмінних ефектів, так і без. Крім стандартного стохастичного методу у формалізмі початкових конфігурацій ми також представляємо точний підхід. Наші числові розрахунки підтверджують, що запропонований метод має таке ж сприятливе лінійне масштабування за часами обчислення, як і нещодавно розроблена схема G1–G2 [Schluenzen et al., Phys. Rev. Lett., 2020, \textbf{124}, 076601]. У той же час підходи НСП і SGW краще масштабуються за розміром, ніж схема G1-G2, що дозволяє застосовувати нерівноважні \emph{GW} розрахунки і для більших систем.    
	
	\keywords квантова динаміка, нерівноважні функції Ґріна, квантові флуктуації, модель Габбарда, \emph{GW} наближення
	
\end{abstract}

\lastpage
\end{document}